\documentclass[11pt,a4paper]{article}
\newcommand{\eref}[1]{(\ref{#1})}

\usepackage[T1]{fontenc}
\usepackage{amsmath}
\usepackage{amsthm}
\usepackage{amsfonts}
\usepackage{amssymb}
\usepackage{graphicx,subfigure,color}
\usepackage{hyperref}
\usepackage[font=footnotesize,labelfont=bf]{caption}
\usepackage{amscd}

\usepackage{enumerate}

\begin{document}
\def\thefootnote{\fnsymbol{footnote}}

\begin{center}
\Large{\textbf{General solution of a cosmological model induced from higher dimensions using a kinematical constraint}} \\[0.5cm]

\large{\"{O}zg\"{u}r Akarsu$^{\rm a}$, Tekin Dereli$^{\rm a}$, Nihan Kat{\i}rc{\i}$^{\rm b}$, Mikhail B. Sheftel$^{\rm b}$}
\\[0.5cm]

\small{
\textit{$^{\rm a}$ Department of Physics, Ko\c{c} University, 34450 Sar{\i}yer, {\.I}stanbul, Turkey}}

\vspace{.2cm}

\small{
\textit{$^{\rm b}$ Department of Physics, Bo\u{g}azi\c{c}i University, 34342 Bebek, {\.I}stanbul, Turkey}}

\end{center}

\vspace{.6cm}

\hrule \vspace{0.3cm}
\noindent \small{\textbf{Abstract}\\
In a recent study  Akarsu and Dereli (Gen. Relativ. Gravit. 45:1211, 2013) discussed the dynamical reduction of a higher dimensional cosmological model which is augmented by a kinematical constraint characterized by a single real parameter,  correlating and controlling the expansion of both the external (physical) and internal spaces. In that paper explicit solutions were found only for the case of three dimensional internal space ($n=3$). Here we derive a general solution of the system using Lie group symmetry properties, in parametric form  for arbitrary number $n=1,2,3,\dots$ of internal dimensions. We also investigate the dynamical reduction of the model as a function of  cosmic time $t$ for various values of $n$  and generate parametric plots to discuss  cosmologically relevant results. }
\\
\noindent
\hrule
\def\thefootnote{\arabic{footnote}}
\setcounter{footnote}{0}

\let\thefootnote\relax\footnote{\textbf{E-Mail:} oakarsu@ku.edu.tr, tdereli@ku.edu.tr, nihan.katirci@boun.edu.tr, mikhail.sheftel@boun.edu.tr}

\def\thefootnote{\arabic{footnote}}
\setcounter{footnote}{0}

\section{Introduction}
\label{Intro}
The unification of all fundamental interactions of nature achieved in higher dimensional space-times, e.g. the formulation of consistent anomaly-free superstring models in ten dimensions \cite{Lidsey00}, provides  strong motivation for considering higher dimensional cosmological models. In such approaches, it is generally assumed that all but four of the space-time dimensions are compactified on an unobservable internal manifold, leaving back an observable (1+3)-dimensional space-time. On the other hand the dynamical reduction of internal dimensions to
unobservable scales, with physical external dimensions
expanding while the internal dimensions contracting, has also been
considered in cosmology (See \cite{ChodosDetweiler80,Freund82,DereliTucker83} for the very first papers). Such models are of interest in cosmology particularly since the dynamics of the internal space, though cannot be observed locally and directly today, could modify the dynamics of the external space in various ways and may contribute and even lead to a dynamically accelerated expansion \cite{Townsend03,Neupane04,Neupane05,Akarsu13b,Akarsu13c}. However, then the evolution of the internal space should be slow enough not to contradict local physics, e.g., the observational constraints on the $(1+3)$-dimensional gravitational coupling that is inversely proportional with the volume of the internal space \cite{Dvali99,Uzan11}.

In a recent study, in contrast to the widely considered contracting internal space scenarios in cosmology, Akarsu and Dereli \cite{AkDe} have considered an alternative dynamical reduction scenario in which both of the external and internal dimensions are assumed to be at comparably small scales during the early stages; yet at later stages of the evolution of the universe the internal dimensions expand at a much slower rate than those of the external space and remain unobservable. The idea was demonstrated on a simple higher dimensional cosmological model that is augmented by a kinematical constraint characterized by a single parameter. Namely, the product of the Hubble parameters of the internal and external spaces is set equal to a real constant that correlates and controls the dynamical evolution of the external (physical) and internal spaces. In this study we utilize the same kinematical constraint for obtaining cosmological solutions not only for $3$-dimensional expanding internal space but for that may be either expanding or contracting internal space with arbitrary number of dimensions, thus generalizing our previous paper \cite{AkDe}.

Let us now proceed with a brief outline of the model introduced in \cite{AkDe}. A minimal extension of the conventional $(1+3)$-dimensional Einstein's field equations to $(1+3+n)$-dimensions is considered:
\begin{equation}
\label{eqn:EFE}
R_{\mu\nu}-\frac{1}{2}Rg_{\mu\nu} =-\kappa T_{\mu\nu},
\end{equation}
where the indices $\mu$ and $\nu$ run through $0,1,2,...,3+n$ and
$g_{\mu\nu}$, $R_{\mu\nu}$ and $R$ are the metric tensor, the Ricci
tensor and the Ricci scalar, respectively, of a $(1+3+n)$-dimensional space-time. $T_{\mu\nu}$ is the effective energy-momentum tensor of matter fields in $(1+3+n)$-dimensions and $\kappa = 8\pi G$ where $G$ is the (positive) gravitational constant that is to be scaled consistently in $(1+3+n)$-dimensions. The space-time is described by the spatially homogenous but not necessarily isotropic
$(1+3+n)$-dimensional synchronous space-time metric that involves a
maximally symmetric three dimensional flat external (physical) space
metric and a compact $n$ dimensional flat internal space metric:
\begin{eqnarray}
\label{eqn:metric} {\rm d}s^2=-{\rm d}t^2+a^2(t) \left ({\rm d}x^{2}+{\rm d}y^{2}+{\rm d}z^{2}\right)+ s^2(t) \left ( {\rm d}\theta_{1}^{2} +...+{\rm d}\theta_{n}^{2}\right),
\end{eqnarray}
where $t$ is the cosmic time, $a(t)$ is the scale factor of the external space and $s(t)$ is the scale factor of the $n=1,2,3,\dots$ dimensional internal space. The energy-momentum tensor of a $(1+3+n)$-dimensional ideal fluid is considered to be homogeneous and isotropic:
\begin{equation}
\label{eqn:EMT}
{T^{\mu}}_{\nu}={\textnormal{diag}}[-\rho,p,p,p,p,...,p],
\end{equation}
where $\rho=\rho(t)$ and $p=p(t)$ are the energy density and pressure of the fluid. The $(1+3+n)$-dimensional Einstein's field equations (\ref{eqn:EFE}) for the space-time described by the metric (\ref{eqn:metric}) in the presence of a co-moving fluid represented by the energy-momentum tensor (\ref{eqn:EMT}) read:
\begin{subequations}
\begin{align}
        3\frac{\dot{a}^2}{a^2}+3n\frac{\dot{a}}{a}\frac{\dot{s}}{s}+\frac{1}{2}n(n-1)\frac{\dot{s}^2}{s^2}&=\kappa\rho, \label{eqn:EFE1} \\
        \frac{\dot{a}^2}{a^2}+2\frac{\ddot{a}}{a}+n\frac{\ddot{s}}{s}+2n\frac{\dot{a}}{a}\frac{\dot{s}}{s}+\frac{1}{2}n(n-1)\frac{\dot{s}^2}{s^2}
&= -\kappa p,  \label{eqn:EFE2}\\
3\frac{\dot{a}^2}{a^2}+3\frac{\ddot{a}}{a}+(n-1)\frac{\ddot{s}}{s}+3(n-1)\frac{\dot{a}}{a}\frac{\dot{s}}{s}
+\frac{1}{2}(n-1)(n-2)\frac{\dot{s}^2}{s^2}
&= -\kappa p, \label{eqn:EFE3}
\end{align}
\end{subequations}
where a dot over a symbol designates derivative with respect to cosmic time $t$. Our system consists of three differential equations (\ref{eqn:EFE1})-(\ref{eqn:EFE3}) satisfied by four unknown functions $a$, $s$, $\rho$, $p$, and hence is under-determined. The Akarsu-Dereli model \cite{AkDe}, on the other hand, is characterized by an additional constraint which determines the system fully by fixing the product of the Hubble parameters of the internal and external spaces to a constant
\begin{equation}
\label{constr}
\frac{\dot{a}}{a}\frac{\dot{s}}{s}=\frac{\lambda}{9}.
\end{equation}
 Accordingly, for an expanding external space, i.e., $\frac{\dot{a}}{a}>0$, the internal space expands for $\lambda>0$, is static for $\lambda=0$ and contracts for $\lambda<0$.
In Ref.\cite{AkDe}, explicit exact solutions of the field equations were given only for the particular case where the number of internal dimensions $n=3$  and a detailed discussion of the model follows for $\lambda>0$.  It was shown that the external space behaves similarly as in the standard model of cosmology, say the $\Lambda$-Cold Dark Matter ($\Lambda$CDM) model \cite{Sahni00}, with the difference that powers of $t$ are not the same; it is $\frac{1}{3}$ in Akarsu-Dereli model  while  $\frac{2}{3}$ in  $\Lambda$CDM. The internal dimensions are found to be expanding but at a much slower rate than those of the external dimensions. Therefore, since all dimensions are assumed to be at a comparable size at some early stage of the universe, when they reach the present time of the universe ($\sim14$ Gyr) the external dimensions have expanded $\sim 10^{60}$ times  while the internal dimensions expand only $\sim 1.5$ times their original size.

In what follows,  we derive the general solution of the above system in parametric form for arbitrary number of internal dimensions $n$.
Although we cannot
write it down explicitly in terms of the cosmic time variable $t$, the general solution of the system in parametric form allows us to discuss various features of the model depending on the number of the internal dimensions and to give parametric plots for demonstrating the dynamics of the universe in $t$.


\section{The General Solution}

We start by subtracting \eref{eqn:EFE2} from \eref{eqn:EFE3} in order to eliminate $p$:
\begin{equation}
\label{firststep}
2\frac{\dot{a}^2}{a^2}+\frac{\ddot{a}}{a}-\frac{\ddot{s}}{s}+(n-3)\frac{\dot{a}}{a}\frac{\dot{s}}{s}
-(n-1)\frac{\dot{s}^2}{s^2}=0.
\end{equation}
We then obtain the scale factors of the external and internal dimensions by solving this equation together with the kinematical constraint \eref{constr}. The energy density and pressure of the higher dimensional fluid are found by putting these back in \eref{eqn:EFE1} and \eref{eqn:EFE2} (or \eref{eqn:EFE3}), respectively. We will first discuss the simplest case $\lambda=0$ and  investigate the case $\lambda\neq0$ afterwards. We note from the kinematical constraint \eref{constr} that in case $\lambda=0$, unless both of the internal and external spaces are static, either the internal dimensions or the external dimensions should be static. Accordingly, \eref{firststep} reduces either to
\begin{equation}
\label{ara1}
2\frac{\dot{a}^2}{a^2}+\frac{\ddot{a}}{a}=0
\end{equation}
for static internal space, i.e. $s={\rm constant}$, or to
\begin{equation}
\label{ara2}
(n-1)\frac{\dot{s}^2}{s^2}+\frac{\ddot{s}}{s}=0
\end{equation}
for static external space, i.e. $a={\rm constant}$. Considering the solution of \eref{ara1} and $s={\rm constant}$ in \eref{eqn:EFE1} and \eref{eqn:EFE2} we obtain
\begin{equation}
a=(c_{0}t+c_{1})^{\frac{1}{3}},\quad s=s_{0}\quad\textnormal{and}\quad p=\rho=\frac{{c_{0}}^2}{3\kappa}(c_{0}t+c_{1})^{-2} \quad (\textnormal{Case I for}\; \lambda=0)
\end{equation}
for static internal space. Similarly considering the solution of \eref{ara2} and $a={\rm constant}$ we obtain
\begin{equation}
\label{2lambda0}
a=a_{0},\quad s=(c_{2}t+c_{3})^{\frac{1}{n}}\quad\textnormal{and}\quad p=\rho=\frac{{c_{2}}^{2}}{2\kappa}\frac{n-1}{n}(c_{2}t+c_{3})^{-2} \quad (\textnormal{Case II for}\; \lambda=0)
\end{equation}
for static external space. It maybe noteworthy that if the external space is static  and there is only one extra dimension ($n=1$), then the universe should be empty.

\medskip

The solution of the system \eref{eqn:EFE1}-\eref{constr} in case $\lambda\neq 0$ is not straightforward in contrast to the case $\lambda=0$. This is because \eref{firststep} cannot be solved explicitly in terms of cosmic time $t$ for arbitrary values of $n$ but only for $n=3$. Now using the kinematical constraint \eref{constr} in \eref{firststep} to eliminate $s$ we arrive at a single second order differential equation with one unknown $a=a(t)$
\begin{equation}\label{ode}
    \frac{\ddot{a}}{a} + \frac{\lambda}{9}\frac{\ddot{a}a}{\dot{a}^2} = - 2\frac{\dot{a}^2}{a^2} + \frac{n\lambda^2}{81}\frac{a^2}{\dot{a}^2} + \frac{(4-n)}{9}\lambda.
\end{equation}
Symmetry group analysis yields only two obvious Lie point symmetries of this equation \eref{ode}, namely translations in $t$, since there is no explicit $t$ in \eref{ode}, and scaling in $a$, since equation \eref{ode} is homogeneous in $a$, i.e. it does not change under the scaling transformation $\tilde{a} = k a$ with constant $k$. According to the theory of Lie, the existence of a two parameter Lie group of point symmetries implies the integrability of the second order ODE in quadratures. The corresponding symmetry generators are
\begin{equation}\label{gen}
    X_1 = \frac{1}{n\lambda} \partial_t ,\quad X_2 = a\partial_a
\end{equation}
where $\partial_t = \partial/\partial_t$ and similarly for $\partial_a$. We could skip the constant factor in $X_1$ but it would complicate calculations at later steps. These two generators commute, $[X_1,X_2] = 0$,  and are linearly independent since
the determinant of the matrix of their components is nonzero $\delta = \left|
\begin{array}{l}
   X_1\\
   X_2
\end{array}
\right| = \left|
\begin{array}{lr}
   1/(n\lambda) & 0\\
   0  & a
\end{array}
\right| = \frac{\displaystyle a}{\displaystyle n\lambda} \ne 0$. Therefore, the abelian symmetry Lie group acts transitively on the representation space with the coordinates $(t, a)$ and we have the case $G_2$I$a$ of the book by H. Stephani \cite{steph}. In this case there exist such canonical variables $\tau$ and $\sigma$, where we consider $\sigma$ as a function of $\tau$, $\sigma=\sigma(\tau)$, which are functions of the original variables $t$ and $a$ such that the symmetry generators \eref{gen} take the normal forms
\begin{equation}
\label{normal}
    X_1 = \partial_\sigma,\qquad X_2 = \partial_\tau.
\end{equation}
Variables $\tau$ and $\sigma$ satisfy the equations, obvious from \eref{gen} and \eref{normal}
\begin{equation}
\label{t}
    X_1(\tau) = \frac{1}{n\lambda}\tau_t = 0,\qquad X_2(\tau) = a\tau_a = 1
\end{equation}
and
\begin{equation}\label{s}
    X_1(\sigma) = \frac{1}{n\lambda}\sigma_t = 1,\qquad X_2(\sigma) = a\sigma_a = 0
\end{equation}
where the letter subscripts denote partial derivatives with respect to corresponding variables. Simplest solutions of equations \eref{t} and \eref{s} together with the inverse transformation from $\tau,\sigma$ to $t, a$ have the form
\begin{equation}\label{ts}
    \tau = \ln{|a|},\quad \sigma = n\lambda t,\qquad t = \frac{\sigma}{n\lambda},\quad a = \varepsilon e^\tau
\end{equation}
where $\varepsilon = \textrm{sign}(a)$. We also need to transform the derivatives $\dot{a}$ and $\ddot{a}$ to $\sigma,\tau$ and to the derivatives $\sigma'={\rm d}\sigma/{\rm d}\tau$, $\sigma''={\rm d}\sigma'/{\rm d}\tau$ of the new unknown $\sigma=\sigma(\tau)$ with respect to the new independent variable $\tau$. (From now on, the primes denote derivatives with respect to $\tau$ while the dots designate derivatives with respect to time $t$.) This is done as follows: $\sigma'={\rm d}\sigma/{\rm d}\tau = n\lambda {\rm d}t/((\dot{a}/a){\rm d}t) = n\lambda a/\dot{a}$. We obtain $\dot{a} = \displaystyle\frac{n\lambda a}{\sigma'} = \frac{n\lambda\varepsilon e^\tau}{\sigma'}$. To transform $\ddot{a}$, we consider $\sigma'' = {\rm d}\sigma'/{\rm d}\tau = n\lambda\displaystyle\frac{{\rm d}(a/\dot{a})}{{\rm d}a/a} = n\lambda\frac{(a/\dot{a})\dot{}}{\dot{a}/a} = n\lambda\frac{a}{\dot{a}^3}(\dot{a}^2 - a\ddot{a})$. Using here the transformations $\dot{a}= \displaystyle\frac{n\lambda\varepsilon e^\tau}{\sigma'}$, $a=\varepsilon e^\tau$ and solving algebraically for $\ddot{a}$ we obtain $\ddot{a} = n^2\lambda^2\varepsilon e^\tau
\left(\displaystyle\frac{1}{\sigma^{\prime\,2}} - \frac{\sigma''}{\sigma^{\prime\,3}}\right)$.

Now we insert the transformed $a, \dot{a}, \ddot{a}$ into equation \eref{ode} to obtain after some arithmetics the equation transformed to the canonical variables $\tau$ and $\sigma=\sigma(\tau)$
\begin{equation}\label{ode_st}
    \sigma'' = \frac{\sigma'}{9n\lambda(\sigma^{\prime\,2} + 9n^2\lambda)}\{243n^3\lambda^2 + 9n(n-3)\lambda \sigma^{\prime\,2} - \sigma^{\prime\,4}\}.
\end{equation}
By construction, this equation does not contain $\sigma$ explicitly because it admits translations in $\sigma$ generated by $X_1 = \partial_\sigma$, so that we can reduce its order by one unit choosing $\sigma'$ for the new unknown: $r=\sigma'(\tau)$ and $\sigma'' = {\rm d}r/{\rm d}\tau$. Since it also admits another symmetry $X_2 = \partial_\tau$, it does not contain explicitly $\tau$ either, so that \eref{ode_st} admits separation of variables $r$ and $\tau$ in the form
\begin{equation}\label{separated}
    - 9n\lambda\,\frac{(r^2 + 9n^2\lambda)\,{\rm d}r}{r\{r^4 - 9n(n-3)\lambda r^2 - 243n^3\lambda^2\}} = {\rm d}\tau
\end{equation}
which we immediately integrate. We split the integral on the left-hand side into two parts $J_1$ and $J_2$, so that the integrated equation \eref{separated}
becomes the first integral
\begin{equation}\label{integr}
   - 9n\lambda J = \tau - \ln{|C_1|} = \ln{\left|\frac{a}{C_1}\right|},\quad J = \frac{1}{2}\,J_1 + \frac{9n^2\lambda}{2}\,J_2
\end{equation}
where
\begin{equation}\label{J}
    J_1 = \int \frac{{\rm d}\psi}{P(\psi)},\quad J_2 = \int \frac{{\rm d}\psi}{\psi P(\psi)}
\end{equation}
with $\psi = r^2$ and $P(\psi) = \psi^2 - 9n(n-3)\lambda\psi - 243n^3\lambda^2$. This polynomial admits the factorization:
$P(\psi) = (\psi - 9n^2\lambda)(\psi + 27n\lambda)$. The expansion of the denominators of $J_1$ and $J_2$ leads to the following results
\begin{eqnarray}
    J_1 &=& \frac{1}{9n(n+3)\lambda}\ln{\left|\frac{\psi - 9n^2\lambda}{\psi + 27n\lambda}\right|},
    \label{J_12} \\
    J_2 &=& \frac{1}{243n^3(n+3)\lambda^2}\ln{\left|\frac{(\psi + 27n\lambda)^n(\psi - 9n^2\lambda)^3}{\psi^{n+3}}\right|}. \nonumber
\end{eqnarray}
Using \eref{J_12} in \eref{integr} and eliminating logarithms, we obtain
\begin{equation}\label{a}
    a = C_1 r^{1/3}|r^2 + 27n\lambda|^{-\frac{(n-3)}{6(n+3)}}|r^2 - 9n^2\lambda|^{-\frac{1}{n+3}}.
\end{equation}
According to the routine, e.g. in \cite{steph}, we had to replace here $a$ by $\varepsilon e^\tau$ and solve algebraically equation \eref{a} for $r={\rm d}\sigma/{\rm d}\tau$ as an explicit function of $\tau$ and then integrate once to get $\sigma=\sigma(\tau)$. Performing the inverse transformation from canonical variables $\tau$ and $\sigma$ in the solution to $t$ and $a$ we will obtain the required dependence $a(t)$. However, such strategy would not work here because for $n\ne 3$ it is impossible to solve explicitly the equation \eref{a} for $r$. Therefore, our approach will be to regard the result \eref{a} as determining $a$ as a function of parameter $r$. Then we need to have also another variable $t$ as a function of $r$. For this purpose let us rewrite our previous relation $r = \sigma' = n\lambda a/({\rm d}a/{\rm d}t)$ in the form ${\rm d}t = \displaystyle \frac{r}{n\lambda}{\rm d}(\ln{|a|})$ with $\ln{|a|}$ calculated from \eref{a} and differentiated afterwards with the following result
\begin{equation}\label{dtau}
    {\rm d}t = \frac{9}{n+3}\left\{\frac{n-3}{r^2+27n\lambda} - \frac{2n}{r^2-9n^2\lambda}\right\}{\rm d}r.
\end{equation}
Integration of this equation yields
\begin{equation}\label{tau}
    t = \frac{9}{n+3}\left\{(n-3)I_1 - 2nI_2\right\},\quad I_1 = \int\frac{{\rm d}r}{r^2+27n\lambda},\quad I_2 = \int\frac{{\rm d}r}{r^2-9n^2\lambda}.
\end{equation}
Calculation of the integrals $I_1, I_2$ depends on the sign of $\lambda$:
\bigskip

\textbf{Case 1:} $\lambda>0$.
\begin{equation}\label{I_pos_la}
    I_1 = \frac{1}{3\sqrt{3n\lambda}}\arctan{\left(\frac{r}{3\sqrt{3n\lambda}}\right)},\quad
    I_2 = \frac{1}{6n\sqrt{\lambda}}\,\ln{\left|\frac{r-3n\sqrt{\lambda}}{r+3n\sqrt{\lambda}}\right|}.
\end{equation}
\bigskip

\textbf{Case 2:} $\lambda= - \mu^2 < 0$.
\begin{equation}\label{I_neg_la}
    I_1 = \frac{1}{6\sqrt{3n}\mu}\,\ln{\left|\frac{r-3\sqrt{3n}\mu}{r+3\sqrt{3n}\mu}\right|},\quad
    I_2 = \frac{1}{3n\mu}\arctan{\left(\frac{r}{3n\mu}\right)}.
\end{equation}
Using these values of the integrals $I_1$ and $I_2$ in \eref{tau} we obtain the following final expressions for $t$ in both cases:

\textbf{Case 1:} $\lambda>0$.
\begin{equation}\label{tau_pos_la}
    t = \frac{\sqrt{3}}{(n+3)\sqrt{n\lambda}}\left\{(n-3)\arctan{\left(\frac{r}{3\sqrt{3n\lambda}}\right)}
    - \sqrt{3n}\ln{\left|\frac{r-3n\sqrt{\lambda}}{r+3n\sqrt{\lambda}}\right|}\right\} + C_2.
\end{equation}
\bigskip

\textbf{Case 2:} $\lambda= - \mu^2 < 0$.
\begin{equation}\label{tau_neg_la}
    t = \frac{\sqrt{3}}{(n+3)\sqrt{n}\mu}\left\{\frac{(n-3)}{2}\ln{\left|\frac{r-3\sqrt{3n}\mu}{r+3\sqrt{3n}\mu}\right|}
    - 2\sqrt{3n}\arctan{\left(\frac{r}{3n\mu}\right)}\right\} + C_2.
\end{equation}
Equation \eref{a} for $a(r)$ together with either \eref{tau_pos_la} for $\lambda>0$ or alternatively \eref{tau_neg_la} for $\lambda < 0$ with $\mu = \sqrt{-\lambda}>0$, which determine $t(r)$,
yields the required general solution of equation \eref{ode} in a parametric form with the parameter $r$. Here $C_1$ and $C_2$ are two arbitrary constants which should be present in the general solution. We note here that, equation \eref{a} on the other hand is valid for both signs of $\lambda$; for $\lambda<0$ we use it with $\lambda=-\mu^2$.

We gave the general solution for the scale factor $a$ in terms of parameter $r$ in \eref{a} and the cosmic time $t$ in terms of parameter $r$ depending on the sign of the constant $\lambda$ in \eref{tau_pos_la} and \eref{tau_neg_la}. Hence, we have now the general solution of \eref{ode} and we further need to obtain the solution of $s$, $\rho$ and $p$ to obtain the general solution of the model determined by equations \eref{eqn:EFE1}-\eref{constr}. We now proceed with determining the scale factor of the internal space $s(t)$ which is in general possible only in a parametric form. We use the kinematical constraint \eref{constr} where $a$ is already known in a parametric form. We need only our previous result $\dot{a}/a = n\lambda/r$ which by \eref{constr} implies $\dot{s}/s =r/(9n)$ or equivalently ${\rm d}(\ln{|s|}) = r{\rm d}t/(9n)$, where we will use equation \eref{dtau} for ${\rm d}t$ which is valid for both signs of $\lambda$. We obtain
\begin{equation}\label{ds}
    {\rm d}(\ln{|s|}) = \frac{1}{n+3}\left\{\frac{(n-3)}{n(r^2+27n\lambda)} - \frac{2}{r^2-9n^2\lambda}\right\}r{\rm d}r
\end{equation}
which by introducing the new variable $\psi=r^2$ is easily integrated in the form
\[\ln{|s|} = \frac{1}{n+3}\left\{\frac{n-3}{2n} \ln{|\psi+27n\lambda|} - \ln{|\psi-9n^2\lambda|}\right\} + \ln{C_3}.\]
We finally obtain $s(t)$ in the parametric form
\begin{equation}\label{s(r)}
    s = C_3|r^2+27n\lambda|^{\frac{n-3}{2n(n+3)}}\; |r^2-9n^2\lambda|^{-\frac{1}{n+3}}
\end{equation}
together with $t(r)$ determined by \eref{tau_pos_la} or \eref{tau_neg_la} depending on the sign of $\lambda$.
To drop the module signs correctly in the equations given in \eref{s(r)} for $s(r)$ and \eref{tau_pos_la}, \eref{tau_neg_la} for $t(r)$, we have again to distinguish different subcases considered above.

Now we use equation \eref{eqn:EFE1} with $a(r)$ and $s(r)$ already determined to find the unknown $\rho$.
\begin{equation}\label{ro}
    \rho = \frac{1}{\kappa}\left(\frac{3n^2\lambda^2}{r^2} + \frac{n\lambda}{3} + \frac{n-1}{162 n}\,r^2\right)
\end{equation}
which again together with $t(r)$ determined by \eref{tau_pos_la} or \eref{tau_neg_la} yields $\rho(t)$ in a parametric form.

\medskip

Finally we will determine $p(t)$ from \eref{eqn:EFE2} (or \eref{eqn:EFE3}) with $a$ and $s$ already found, so that we use $\dot{a}/a = n\lambda/r$,
$\ddot{a}/a = (\dot{a}/a)\dot{} + (\dot{a}/a)^2$ where $(\dot{a}/a)\dot{} = - (n\lambda/r^2){\rm d}r/{\rm d}t$ and ${\rm d}r/{\rm d}t$ is determined by \eref{dtau} as a function of $r$  as
\begin{equation}\label{dr/dtau}
    \frac{{\rm d}r}{{\rm d}t} = - \frac{(r^2+27n\lambda)(r^2-9n^2\lambda)}{9(r^2+9n^2\lambda)}.
\end{equation}
This implies the result
\[\frac{\ddot{a}}{a} = \frac{n\lambda}{9r^2(r^2+9n^2\lambda)}\,[r^4 + 9n(4-n)\lambda r^2 - 162 n^3\lambda^2].\]
We determine $\ddot{s}/s$ in a similar way, using $\dot{s}/s = r/(9n)$, $\ddot{s}/s = (\dot{s}/s)\dot{} + (\dot{s}/s)^2$ and $(\dot{s}/s)\dot{} = (1/(9n)){\rm d}r/{\rm d}t$ with ${\rm d}r/{\rm d}t$ again determined by \eref{dr/dtau}. We obtain
\[\frac{\ddot{s}}{s} = \frac{(1-n)r^4 + 9n^2(n-2)\lambda r^2 + 243n^4\lambda^2}{81n^2(r^2+9n^2\lambda)}.\]
Using all these results \eref{eqn:EFE2} (or \eref{eqn:EFE3}) after some arithmetical simplifications we obtain
\begin{eqnarray}
p=\frac{1}{\kappa}\left\{\frac{3n^2\lambda^2}{r^2}-\frac{(n+4)\lambda}{3}-\frac{(27\lambda n^2-r^2)(n-1)-216n\lambda}{162n(9n^2\lambda+r^2)}r^2\right\}
    \label{p}
\end{eqnarray}
which yields $p(t)$ in a parametric form together with $t(r)$ determined by \eref{tau_pos_la} or \eref{tau_neg_la}. Equation of state parameter of the $(1+3+n)$-dimensional fluid defined as $w=\frac{p}{\rho}$ then can also be given in parametric form using \eqref{p} and \eqref{ro}.

\newpage

\section{Cosmological Models}
Up to this point,  the general solution of our model is obtained in parametric form in terms of $\lambda$, $n$ and a new variable $r$.
 We determined the scale factors of the external space $a$ in \eqref{a} and  of the internal space $s$ in \eqref{s(r)}, as well as the energy density $\rho$ in \eqref{ro} and pressure $p$ in \eqref{p} of the higher dimensional effective fluid.
 Unfortunately it is not possible to write down an analytic expression for $r$  as a function of the physically relevant cosmic time variable  $t$. Yet numerical techniques can be used to generate parametric plots of physical quantities such as the Hubble and deceleration parameters of both external and internal spaces as functions of cosmic time $t$. In order to do that the sign of $\lambda$ should be taken into account and different ranges of $r$ over which the solutions are valid must be determined.
 Accordingly, in what follows we distinguish between different cases depending on the sign of $\lambda$ and the ranges of $r$ over which the general solution is valid. We provide in each case parametric plots given in terms of $t$ for various values of $n=1,2,3,\dots$, thus  demonstrating the cosmological consequences of our model focusing on the behavior of the (physical, three dimensional) external  space.

\subsection{Case $\lambda>0$}
$\lambda>0$ is the case in which the external and internal spaces behave in the same way, namely, as the external space expands/contracts the internal space expands/contracts too. In this case, the cosmic time $t$ is given as
\begin{equation}\label{tau1a}
    t = \frac{\sqrt{3}}{(n+3)\sqrt{n\lambda}}\left\{(n-3)\arctan{\left(\frac{r}{3\sqrt{3n\lambda}}\right)}
    - \sqrt{3n}\ln{\left|\frac{r-3n\sqrt{\lambda}}{r+3n\sqrt{\lambda}}\right|}\right\} + C_2.
\end{equation}
The scale factor of the external space is given as
\begin{equation}\label{a1a}
  a = C_1 r^{1/3}(r^2 + 27n\lambda)^{-\frac{(n-3)}{6(n+3)}}|r^2 - 9n^2\lambda|^{-\frac{1}{n+3}},
\end{equation}
which yields the following Hubble and deceleration parameters
\begin{equation}\label{a1ahq}
H_{a}=\frac{\dot{a}}{a}=n\frac{\lambda}{r} \quad\textnormal{and}\quad q_{a}=-\frac{\ddot{a}a}{\dot{a}^2}=\frac{(r^2+27n\lambda)(9n^2\lambda-r^2)}{9n\lambda(9n^2\lambda+r^2)}-1,
\end{equation}
respectively. The scale factor of the internal dimensions is obtained as
\begin{equation}\label{a1b}
s = C_3(r^2+27n\lambda)^{\frac{n-3}{2n(3+n)}}   |r^2-9n^2\lambda|^{-\frac{1}{3+n}},
\end{equation}
which yields the following Hubble and deceleration parameters
\begin{equation}\label{a1bhq}
H_{s}=\frac{\dot{s}}{s}=\frac{r}{9n} \quad\textnormal{and}\quad q_{s}=-\frac{\ddot{s}s}{\dot{s}^2}=-\frac{n(r^2+27n\lambda)(9n^2\lambda-r^2)}{r^2(9n^2\lambda+r^2)}-1,
\end{equation}
respectively. We note that there are two different sets of solutions according to the sign of $r^2-9n^2\lambda$. We note also that $t'=0$ has real solution neither for the case $r^2-9n^2\lambda>0$ nor for the case $r^2-9n^2\lambda<0$, which implies that we will not need to further concern with the ranges once we consider one of these two ranges provided that $\lambda>0$. It is evident from \eref{a1ahq} that the positive values of $r$ should be considered for expanding universe solutions in this case $\lambda>0$.

\subsubsection{Subcase $r^2-9n^2\lambda < 0.$}
\label{solution221}
\begin{figure}[h!]
     \begin{center}
        \subfigure[]{%
            \label{fig:sf1b}
            \includegraphics[width=0.48\textwidth]{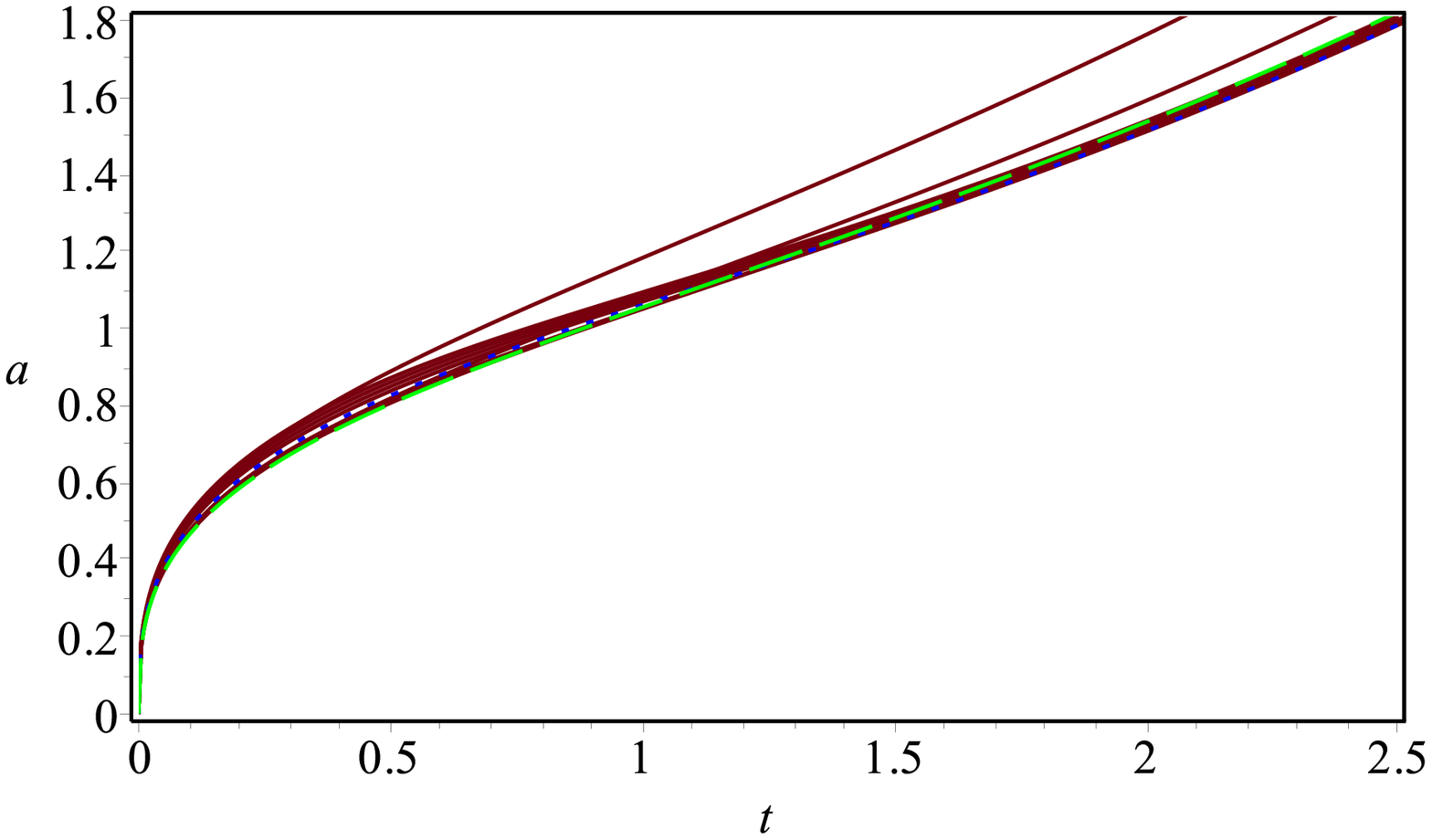}
        }%
        \subfigure[]{%
           \label{fig:q1b}
           \includegraphics[width=0.48\textwidth]{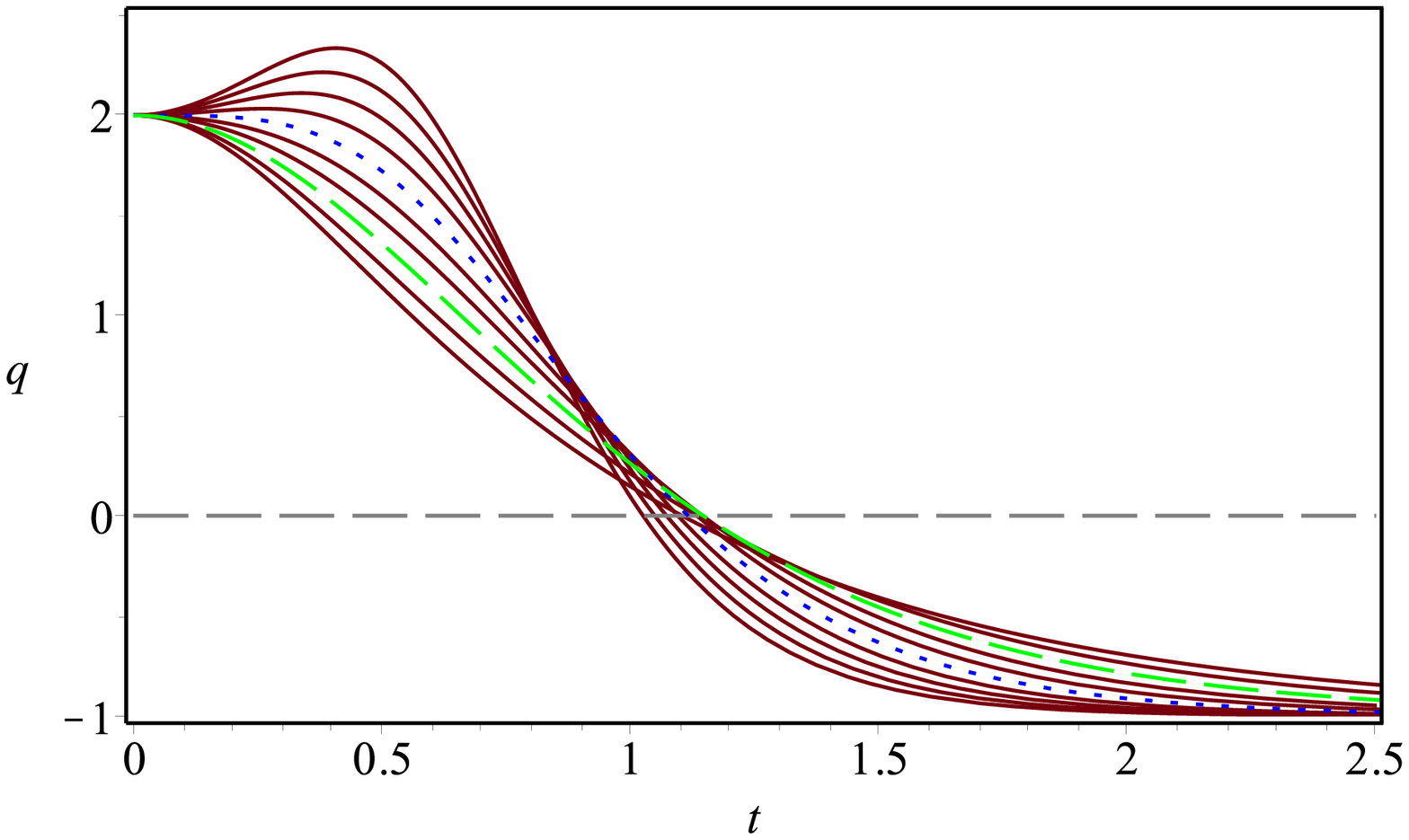}
        }
         \subfigure[]{%
           \label{fig:rho1b}
           \includegraphics[width=0.48\textwidth]{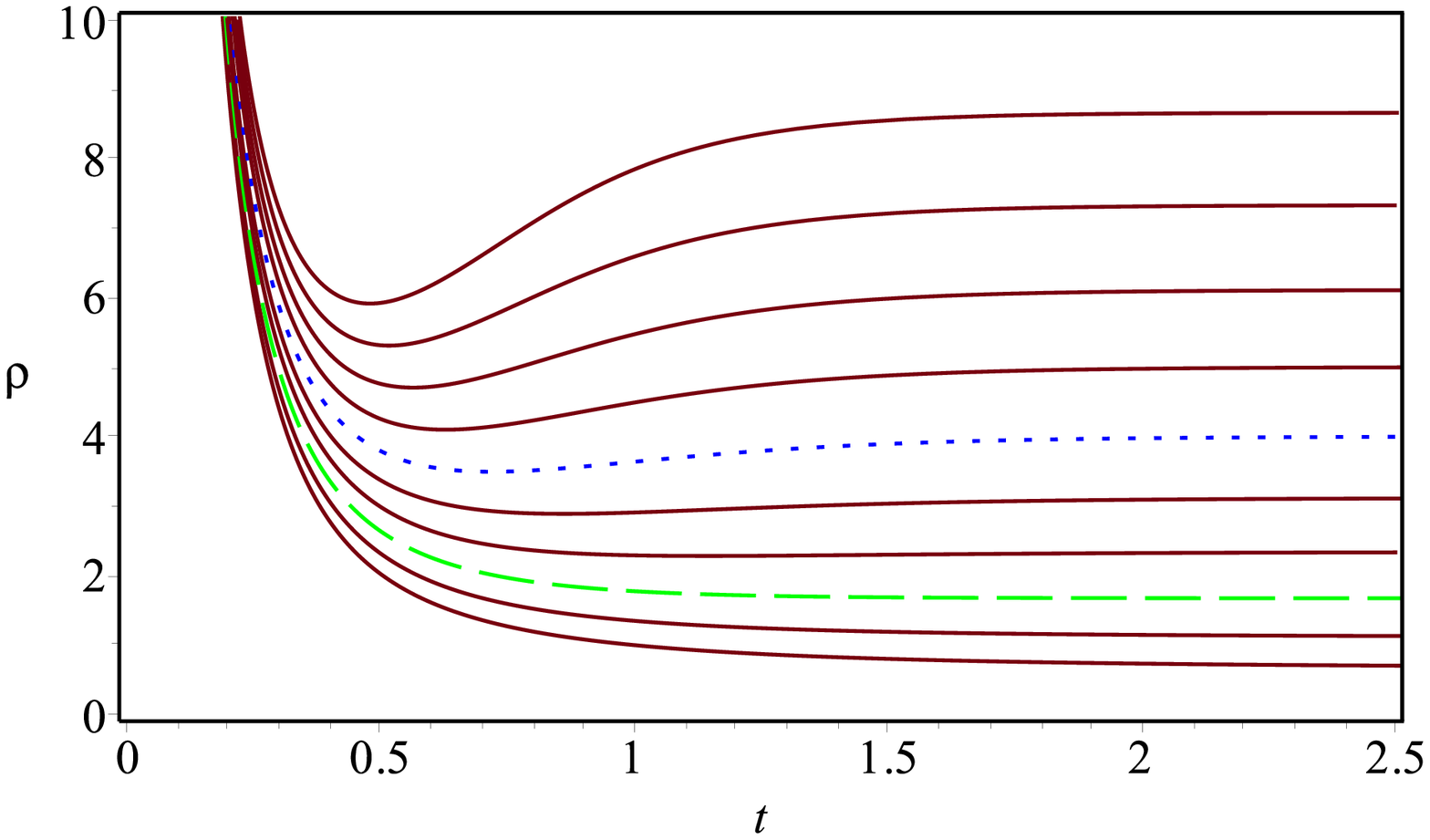}
        }%
         \subfigure[]{%
           \label{fig:eos1b}
           \includegraphics[width=0.48\textwidth]{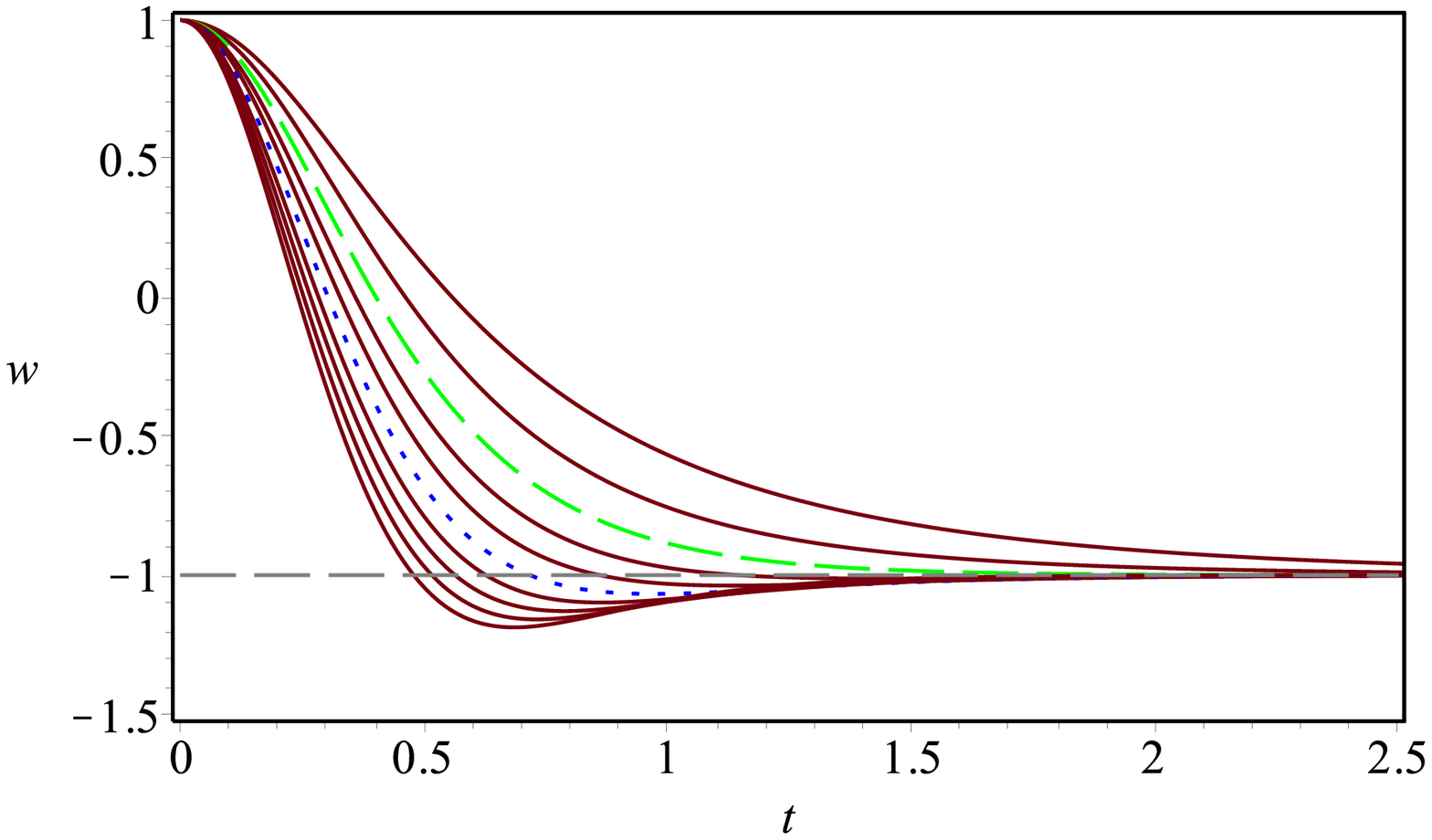}
        }

    \end{center}
	\caption{The evolution of some parameters of the model in cosmic time $t$ for $(4+n)$-dimensions. The plots are given for $n=1$ to $n=10$ by choosing $\lambda=1$. The curves are in an order such that the dashed curves correspond to the case $n=3$ and dotted curves correspond to the case $n=6$. (a) The scale factor of the external dimensions, $a$. (b) The deceleration parameter of the external dimensions, $q_{a}$. (c) The energy density of the higher dimensional fluid, $\rho$. (d) Equation of state (EoS) parameter of the higher dimensional fluid, $w$.}
\end{figure}

This subcase implies $r\in (-3n\sqrt{\lambda},3n\sqrt{\lambda})$. Because we are interested in expanding external space then we consider only the range $0 \leq r \leq 3n\sqrt{\lambda}$. We note that $a\rightarrow 0$, $H_{a}\rightarrow\infty$, $q_{a}\rightarrow 2$ and $s\rightarrow {\rm s_{\rm min}}={\rm const.}$, $H_{s}\rightarrow 0$, $q_{s}\rightarrow-\infty$ as $r \rightarrow 0$ while $a\rightarrow\infty$ and $s\rightarrow\infty$ such that $H_{a}\rightarrow H_{s}\rightarrow \frac{\sqrt{\lambda}}{3}$ and $q_{a}\rightarrow q_{s}\rightarrow -1$ as $r \rightarrow 3n\sqrt{\lambda}$. We note further that $t\rightarrow C_2$ as $r\rightarrow0$ and $t\rightarrow\infty$ as $r \rightarrow 3n\sqrt{\lambda}$ and also that the cosmic time $t(r)$ evolves monotonically between these two limits. Hence external dimensions start expanding from a zero size while internal dimensions start expanding with a non-zero size at $t=0$ (we set $C_2=0$). External dimensions expand always with a higher rate than the external dimensions. All the dimensions, on the other hand, approach the exponential expansion with a same power as $t\rightarrow\infty$. However, we note that their evolution trajectories are dependent on the number of the internal dimensions $n$. To demonstrate to behavior of the model in cosmic time $t$ we presented the parametric plots of the scale factor in Fig. \ref{fig:sf1b} and the deceleration parameter in Fig. \ref{fig:q1b} of the external dimensions versus cosmic time $t$ for $n=1$ to $n=10$. The dashed curves in the figures represent the case $n=3$ whose explicit solution in terms of cosmic time $t$ is available and given below in \eref{1atau} and the dotted curves represent the case $n=6$. Hence, one may have an idea about the behavior of the model depending on the number of internal dimensions $n$ by checking the explicit functions of cosmic time $t$ given in equation \eref{1atau} for $n=3$.
We note a difference in the evolution of the deceleration parameter in accordance to whether the universe has more than 10 dimensions or not: In the range we consider in this solution, $q'=0$ has only one real root ($r_{1}=0$) if $n\leq6$ while it has two real roots ($r_{1}=0$ and $r_{2}=3\sqrt{-n^2\lambda+n\lambda\sqrt{2n^2-6n}}$) if $n>6$. Accordingly, as can also be seen in Fig. \ref{fig:q1b}, if the number of the internal dimensions are higher than $6$, then the deceleration parameter first increases to a certain value and then evolves to $-1$, while in the cases for $n\leq 6$ the deceleration parameter evolves monotonically to $-1$ as the universe expands. The energy density of the higher dimensional fluid is always positive and evolves from infinitely large values $t=0$ to a non-zero constant $\rho\rightarrow\lambda(n+3)(n+2)/18$ as $t\rightarrow\infty$. The EoS parameter of the fluid $w$, on the other hand, starts with the value of $w=1$ at $t=0$ and approaches $w\rightarrow-1$, i.e. cosmological constant/vacuum energy, as $t\rightarrow\infty$.  In Fig. \ref{fig:rho1b} and Fig. \ref{fig:eos1b}, we present the parametric plots of the cosmic time $t$ evolution of the energy density and EoS parameter of the higher dimensional fluid for $n=1$ to $n=10$.

\subsubsection*{The particular case $n=3$}
In the particular case $n=3$ we can obtain the explicit solution of the model in terms of the cosmic time $t$. In this case we have $r^2-81\lambda < 0$, i.e. $r\in(-9\sqrt{\lambda}, 9\sqrt{\lambda})$. Now substituting $n=3$ in \eref{tau1a} we obtain cosmic time $t$ as
\begin{equation}\label{tau1b_n3}
    t = - \frac{1}{2\sqrt{\lambda}}\ln{\left(\frac{9\sqrt{\lambda}-r}{9\sqrt{\lambda}+r}\right)} + C_2
\end{equation}
which is solved for $r(t)$ as follows
\begin{equation}\label{r1b}
    r = 9\sqrt{\lambda}\tanh{[\sqrt{\lambda}(t - C_2)]}.
\end{equation}
Using \eref{r1b} for $r(t)$ and substituting $n=3$ in equations \eref{a1a}-\eref{a1bhq} we obtain
\begin{equation}\label{1atau}
    a = C_1\sinh^{1/3}{[\sqrt{\lambda}(t - C_2)]}, \quad H_{a}=\frac{\sqrt{\lambda}}{3}\coth[\sqrt{\lambda}(t - C_2)] \quad\textnormal{and}\quad q_{a}=-1+3\, {\rm sech}^{2}[\sqrt{\lambda}(t - C_2)]
\end{equation}
for the external space, and
\begin{equation}
    s = C_3\cosh^{1/3}{[\sqrt{\lambda}(t - C_2)]}, \quad H_{s}=\frac{\sqrt{\lambda}}{3}\tanh[\sqrt{\lambda}(t - C_2)] \quad\textnormal{and}\quad q_{s}=-1-3\, {\rm cosech}^{2}[\sqrt{\lambda}(t - C_2)]
\end{equation}
for the internal space. We note that this is the solution investigated in detail in \cite{AkDe}, and hence one may see reference \cite{AkDe} for a comprehensive discussion on the cosmological aspects of this solution.

\subsubsection{Subcase $r^2-9n^2\lambda > 0$}
\begin{figure}[h!]
     \begin{center}
        \subfigure[]{%
            \label{fig:sf1a}
            \includegraphics[width=0.48\textwidth]{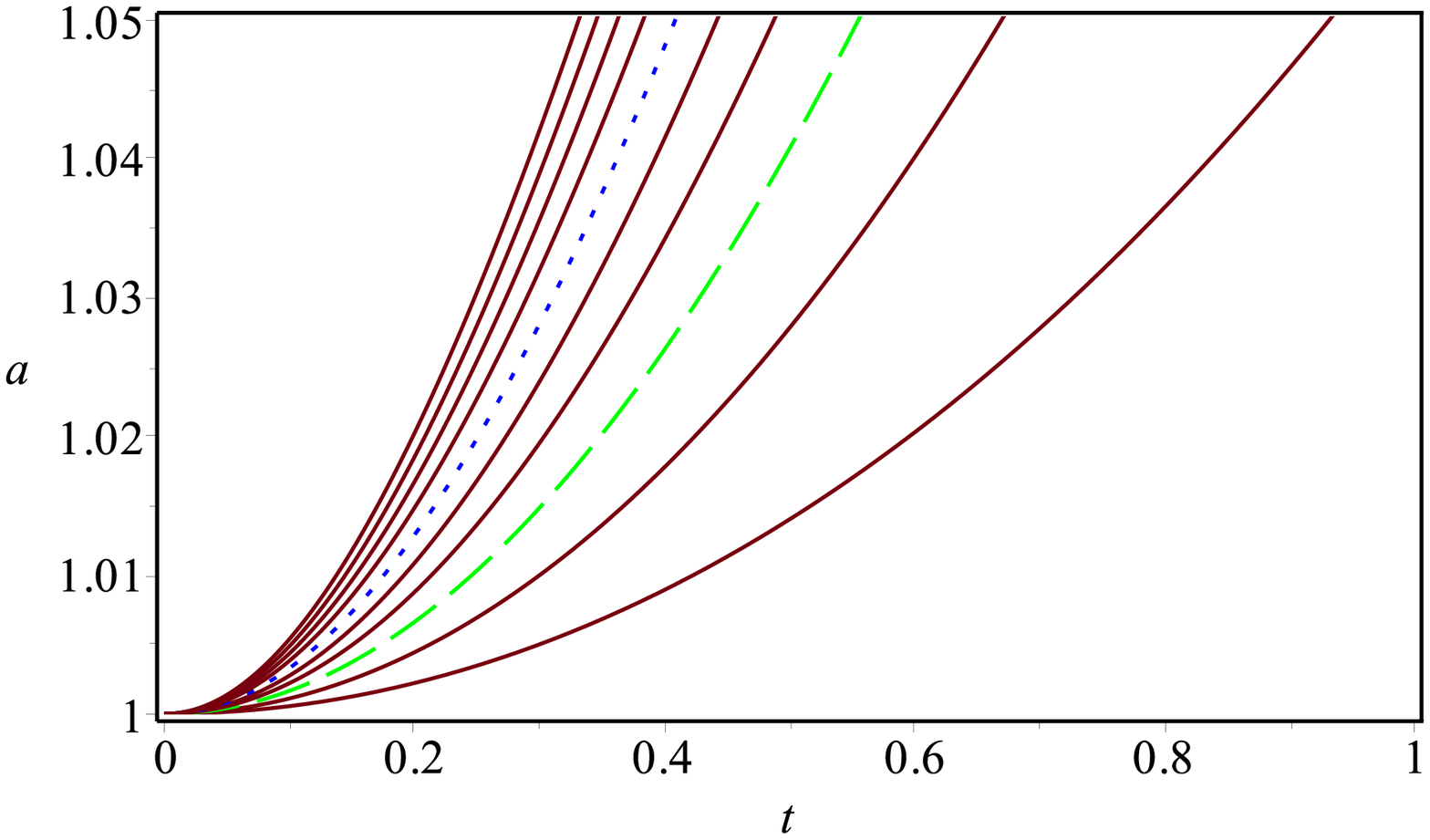}
        }%
        \subfigure[]{%
           \label{fig:q1a}
           \includegraphics[width=0.48\textwidth]{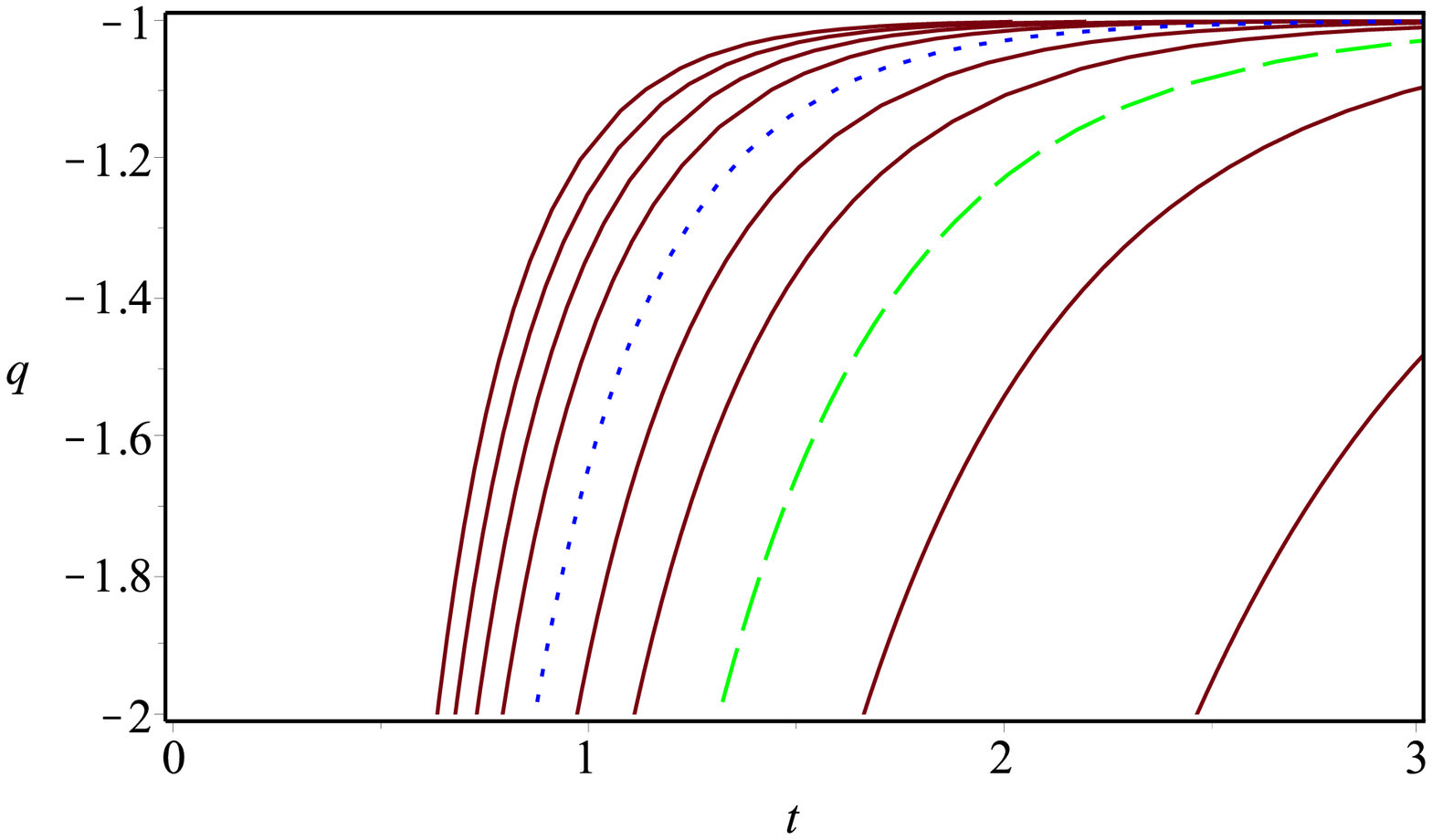}
        }
  \subfigure[]{%
           \label{fig:rho1a}
           \includegraphics[width=0.48\textwidth]{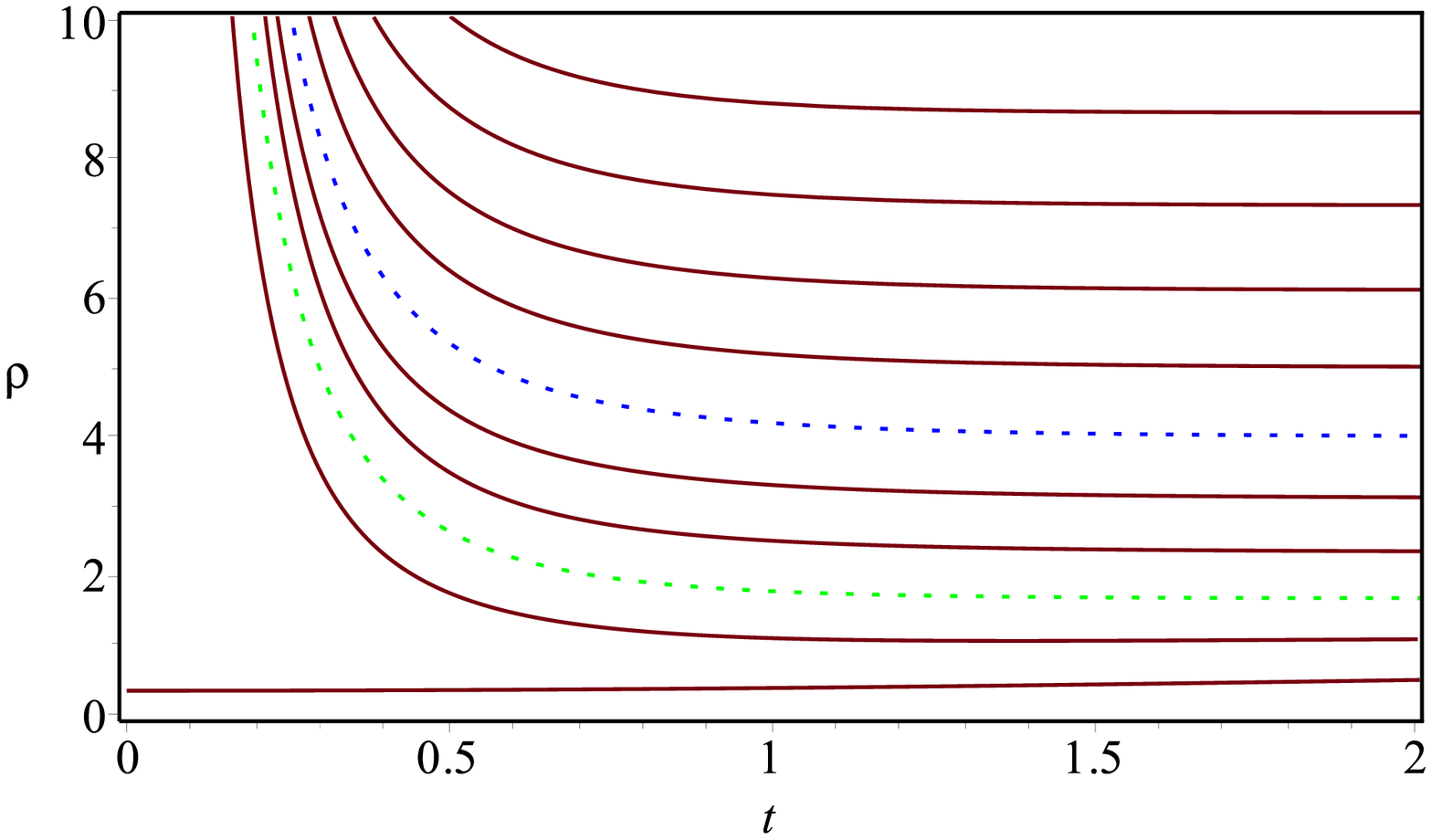}
        }%
         \subfigure[]{%
           \label{fig:eos1a}
           \includegraphics[width=0.48\textwidth]{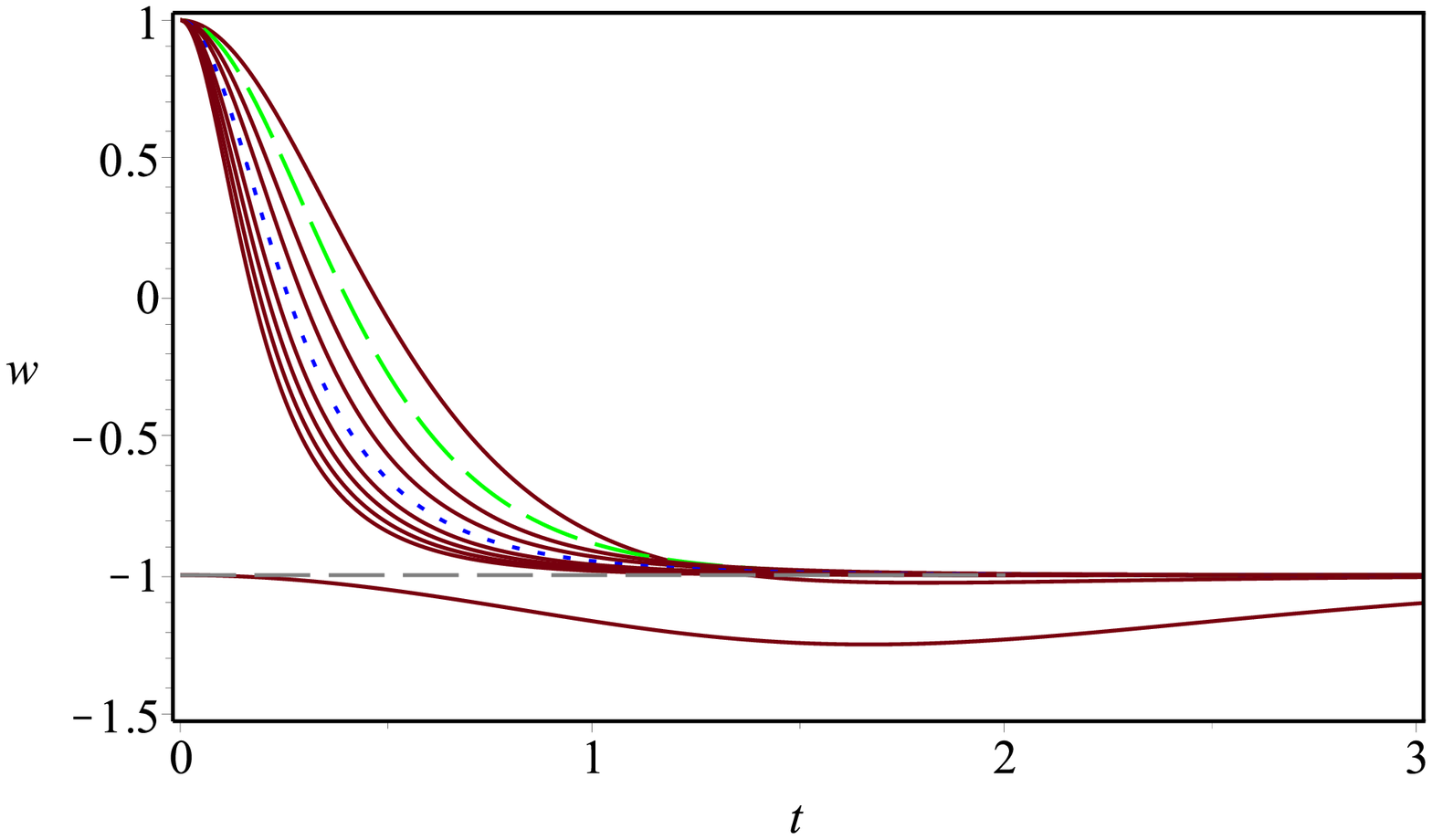}
        }

    \end{center}
	\label{fig:statefinders}
	\caption{The evolution of some parameters of the model in cosmic time $t$ for $(4+n)$-dimensions. The plots are given from $n=1$ to $n=10$ by choosing $\lambda=1$. The curves are in an order such that the dashed curves correspond to the case $n=3$ and dotted curves correspond to the case $n=6$. (a) The scale factor of the external dimensions, $a$. (b) The deceleration parameter of the external dimensions, $q_{a}$. (c) The energy density of the higher dimensional fluid, $\rho$. (d) Equation of state (EoS) parameter of the higher dimensional fluid, $w$.}
\end{figure}

This subcase implies $r\in\{(-\infty,-3n\sqrt{\lambda})\bigcup (3n\sqrt{\lambda},+\infty)\}$ and the expanding external space solution we want to consider is possible in the range $3n\sqrt{\lambda}\leq r \leq \infty$.  We note that $a\rightarrow a_{\rm min}={\rm const.}$, $H_{a}\rightarrow 0$, $q_{a}\rightarrow -\infty$ and $s\rightarrow 0$, $H_{s}\rightarrow \infty$, $q_{s}\rightarrow 2$ as $r \rightarrow \infty$ while $a\rightarrow\infty$ and $s\rightarrow\infty$ such that $H_{a}\rightarrow H_{s}\rightarrow \frac{\sqrt{\lambda}}{3}$ and $q_{a}\rightarrow q_{s}\rightarrow -1$ as $r \rightarrow 3n\sqrt{\lambda}$. We note further that $t\rightarrow C_2+\frac{\pi}{2}\frac{n-3}{n+3}\sqrt{\frac{3}{n\lambda}}$ as $r\rightarrow\infty$ and $t\rightarrow\infty$ as $r\rightarrow 3n\sqrt{\lambda}$ and also that $t(r)$ evolves monotonically between these two limits as in the case $0 \leq r \leq 3n\sqrt{\lambda}$ discussed in section \ref{solution221}. On the other hand, in contrast to the case $0 \leq r \leq 3n\sqrt{\lambda}$, in this case $q_{a}$ is monotonic for all values of $n$. We present the parametric plots of the scale factor in Fig. \ref{fig:sf1a} and the deceleration parameter in Fig. \ref{fig:q1a} of the external dimensions versus cosmic time $t$ for $n=1$ to $n=10$. The dashed green curves in the figures represent the case $n=3$ whose explicit solution in terms of cosmic time $t$ is available and given below in \eref{1atau2} and the dotted curves represent the case $n=6$. Hence, one may have an idea about the behavior of the model depending on the number of internal dimensions $n$ by checking the explicit functions given in equation \eref{1atau2} for $n=3$.
When we consider the higher dimensional fluid in this solution, we note that the case $n=1$ exhibits a qualitatively different behavior than the cases $n\geq2$. For all values of $n$, the energy density of the higher dimensional fluid $\rho$ is always positive and $\rho\rightarrow \lambda(n+3)(n+2)/(18\kappa)$ and $w\rightarrow-1$ as $r\rightarrow 3n\sqrt{\lambda}$ (viz. $t\rightarrow\infty$). However, at the limit $r\rightarrow\infty$ (viz. when the external space starts to expand), $\rho\rightarrow\infty$ and $w\rightarrow1$ for the cases $n\geq2$, while $\rho\rightarrow\frac{\lambda}{(3\kappa)}$ and $w\rightarrow-1$ for the case $n=1$. Accordingly, in the case $n=1$ the higher dimensional fluid yields the form of vacuum energy both at the beginning of the expansion of the external space and in the infinite future, though it starts with an energy density equal to $\frac{2}{(3\kappa)}\lambda$ and ends up with an energy density equal to $\frac{1}{(3\kappa)}\lambda$. In Fig. \ref{fig:rho1a} and Fig. \ref{fig:eos1a}, we present the parametric plots of the cosmic time $t$ evolution of the energy density and EoS parameter of the higher dimensional fluid for $n=1$ to $n=10$. We should, however, note that this solution cannot be considered as a viable higher dimensional cosmological model for any values of $n$ due to two obvious reasons: (i) The internal dimensions expand always faster than the external dimensions. (ii) The deceleration parameter of the external dimensions is always less than $-1$.

\subsubsection*{The particular case $n=3$}
In the particular case $n=3$ we can obtain the explicit solution of the model in terms of the cosmic time $t$. In this case we have $r^2-81\lambda > 0$, i.e. $r\in\{(-\infty,-9\sqrt{\lambda})\bigcup (9\sqrt{\lambda},+\infty)\}$, and the case $r\in(9\sqrt{\lambda},+\infty)$ corresponds to the expanding external space solution. Substituting $n=3$ in \eref{tau1a} we obtain cosmic time $t$ as
\begin{equation}\label{tau1a_n3}
   t = - \frac{1}{2\sqrt{\lambda}}\ln{\left(\frac{r-9\sqrt{\lambda}}{r+9\sqrt{\lambda}}\right)} + C_2
\end{equation}
which is solved for $r(t)$ as follows
\begin{equation}\label{r1a}
    r = 9\sqrt{\lambda}\coth{[\sqrt{\lambda}(t - C_2)]}.
\end{equation}
Using \eref{r1a} for $r(t)$ and substituting $n=3$ in equations \eref{a1a}-\eref{a1bhq} we obtain
\begin{equation}\label{1atau2}
    a = C_1\cosh^{1/3}{[\sqrt{\lambda}(t - C_2)]}, \quad H_{a}=\frac{\sqrt{\lambda}}{3}\tanh[\sqrt{\lambda}(t - C_2)] \quad\textnormal{and}\quad q_{a}=-1-3\, {\rm cosech}^{2}[\sqrt{\lambda}(t - C_2)]
\end{equation}
for the external space and
\begin{equation}
    s = C_3\sinh^{1/3}{[\sqrt{\lambda}(t - C_2)]}, \quad H_{s}=\frac{\sqrt{\lambda}}{3}\coth[\sqrt{\lambda}(t - C_2)] \quad\textnormal{and}\quad q_{s}=-1+3\, {\rm sech}^{2}[\sqrt{\lambda}(t - C_2)]
\end{equation}
for the internal space.

\subsection{Case $\lambda= - \mu^2 < 0$}

$\lambda= - \mu^2 < 0$ is the case in which the external and internal spaces behave in the opposite ways; namely, as the external space expands/contracts the internal space contracts/expands. In this case, the cosmic time $t$ is given as follows:
\begin{equation}
    t = \frac{\sqrt{3}}{(n+3)\sqrt{n}\mu}\left\{\frac{(n-3)}{2}\ln{\left|\frac{r-3\sqrt{3n}\mu}{r+3\sqrt{3n}\mu}\right|}
    - 2\sqrt{3n}\arctan{\left(\frac{r}{3n\mu}\right)}\right\} + C_2.
\end{equation}
The scale factor of the external space is given as
\begin{equation}
  a = C_1 r^{1/3}|r^2 - 27n\mu^2|^{-\frac{(n-3)}{6(n+3)}}(r^2 + 9n^2\mu^2)^{-\frac{1}{n+3}},
\end{equation}
which yields the following Hubble and deceleration parameters
\begin{equation}
\label{sonha}
H_{a}=-n\frac{\mu^2}{r} \quad\textnormal{and}\quad q_{a}=\frac{(27n\mu^2-r^2)(9n^2\mu^2+r^2)}{9n\mu^2(9n^2\mu^2-r^2)}-1,
\end{equation}
respectively. The scale factor of the internal dimensions is obtained as
\begin{equation}
s = C_3 |r^2 - 27n\mu^2|^{-\frac{(n-3)}{2n(n+3)}}(r^2 + 9n^2\mu^2)^{-\frac{1}{n+3}},
\end{equation}
which yields the following Hubble and deceleration parameters
\begin{equation}
\label{sonhs}
H_{s}=\frac{r}{9n} \quad\textnormal{and}\quad q_{s}=\frac{n(27n\mu^2-r^2)(9n^2\mu^2+r^2)}{r^2(9n^2\mu^2-r^2)}-1.
\end{equation}
We note that, in contrast to the case for $\lambda>0$ and $n=3$, here in this solution for $\lambda=-\mu^2<0$ and $n=3$ the solution becomes free from terms involving absolute value, which implies that the case $n=3$ is determined uniquely regardless of the sign of $r^2-27n\mu^2$. The cases $n\neq 3$, on the other hand, should be treated by considering the sign of $r^2-27n\mu^2$ as in the case for $\lambda>0$, though it is not enough. We note that $t'=0$ has two real solutions; $r=3n\mu$ for positive values of $r$ and $r=-3n\mu$ for negative values of $r$. This implies that, once the sign of $r$ is chosen for obtaining solution, $t(r)$ is not monotonic, namely it has one turning point and hence there could be two different branches of the solutions for a chosen sign of $r$. If we give the complete list, in this case $\lambda=-\mu^2<0$, there are the following 
solutions differing
in the ranges of parameter $r$ and the number of internal dimensions $n$:
\begin{enumerate}
\item
If $n=3$, then either  $r\in (-\infty,0)$ or $r\in (0,+\infty)$.
\item
$r^2 - 27n\mu^2 < 0$:
\begin{enumerate}
\item
For $n=1$ and $n=2$, with the ranges $r\in (-3n\mu,0)$ or $r\in (0,3n\mu)$.
\item
For $n=1$ and $n=2$, with the ranges $r\in (-3\sqrt{3n}\mu,-3n\mu)$ or $r\in (3n\mu,3\sqrt{3n}\mu)$.
\item
For $n\geq4$, with the ranges $r\in (-3\sqrt{3n}\mu,0)$ or $r\in (0,3\sqrt{3n}\mu)$.
\end{enumerate}
\item
$r^2 - 27n\mu^2 > 0$
\begin{enumerate}
\item
For $n=1$ and $n=2$, with the ranges $r\in (-\infty,-3\sqrt{3n}\mu)$ or $r\in (3\sqrt{3n}\mu,+\infty)$
\item
For $n\geq4$, with the ranges $r\in (-3n\mu,-3\sqrt{3n}\mu)$ or $r\in (3\sqrt{3n}\mu,3n\mu)$.
\item
For $n\geq4$, with the ranges $r\in (-\infty,3n\mu)$ or $r\in (3n\mu,+\infty)$.
\end{enumerate}
\end{enumerate}
\bigskip

In contrast to the case $\lambda>0$ where the energy density is always positive, the energy density in this case is given as follows
\begin{equation}
\label{romu}
    \rho = \frac{1}{\kappa}\left(\frac{3n^2\mu^4}{r^2} - \frac{n\mu^2}{3} + \frac{n-1}{162 n}\,r^2\right),
\end{equation}
and, it should be further investigated since it can obtain either negative or positive values:
\begin{equation}
\rho<0\quad\textnormal{for}\quad 9n\mu^2\; \frac{3n-\sqrt{3n^2+6n}}{n-1}<r^2<9n\mu^2\; \frac{3n+\sqrt{3n^2+6n}}{n-1}.
\end{equation}
Its first and second derivatives with respect to parameter $r$ as follows:
\begin{equation}
\kappa\rho'=-\frac{6n^2\mu^4}{r^3}+\frac{1}{81}\frac{(n-1)r}{n} \quad\textnormal{and}\quad \kappa\rho''=\frac{18n^2\mu^4}{r^4}+\frac{1}{81}\frac{(n-1)}{n}.
\end{equation}
Accordingly we have
\begin{equation}
\label{critic}
\rho'=0\quad\textnormal{at}\quad r_{\rm c}=\pm 3\mu 6^{1/4} \frac{n^{3/4}}{(n-1)^{1/4}}.
\end{equation}
Using $r_{c}$ in the second derivative we find that
\begin{equation}
 \frac{{\rm d}^2 \rho}{{\rm d}r^2}\bigg\vert_{r=r_{\rm c}}=\frac{4}{81\kappa}\left(1-\frac{1}{n}\right),
\end{equation}
which is always positive since $n\geq1$. Hence, the energy density of the higher dimensional fluid reaches a negative minimum as
\begin{equation}
\label{rhominaddition}
\rho_{\min}=-\frac{n\mu^2}{3\kappa}\left( 1-\sqrt{\frac{2}{3}-\frac{2}{3n}}  \right)<0,
\end{equation}
as long as $r_{\rm c}$ is covered by the range of $r$ in the solution under consideration. Substituting $r=r_{\rm c}$ from \eref{critic} in \eref{p}, we find further that the pressure of the higher dimensional fluid becomes
\begin{equation}
p=-\rho_{\min}\quad\textnormal{at}\quad r=r_{\rm c}.
\end{equation}
We note that the energy density of the higher dimensional fluid takes negative values at $r=r_{\rm c}$ but its EoS takes the form of vacuum energy at $r=r_{\rm c}$. Therefore the minimum of the energy density of the fluid can be shifted to zero by adding a negative cosmological constant $\Lambda=\frac{\rho_{\rm min}}{\kappa}<0$ into the model.

\subsubsection{The particular case $n=3$}

As we mentioned above, there is no restriction on the range of $r$ for the particular case $n=3$ and it can be chosen either as  $r\in (-\infty,0)$ or as $r\in (0,+\infty)$. In this case the cosmic time $t$ is given by
\begin{equation}\label{2tau n3}
    t = - \frac{1}{\mu}\arctan{\left(\frac{r}{9\mu}\right)} + C_2,
\end{equation}
which implies
\begin{equation}\label{2r_n3}
    r = - 9\mu\tan{[\mu(C_2 - t)]}.
\end{equation}
The scale factor on the other hand reduces to
\begin{equation}\label{2a_n3}
    a = C_1r^{1/3}(r^2 + 81\mu^2)^{-1/6}.
\end{equation}
Using $r$ from \eref{2r_n3} in \eref{2a_n3} and redefining $C_1$ as $-C_1$, we obtain the explicit solution $a(t)$ in the form
\begin{equation}
\label{eqn:sin1}
    a = C_1\sin^{1/3}{[\mu(t-C_{2})]}
\end{equation}
which yields
\begin{equation}
\label{eqn:sin2}
H_{a}=\frac{\mu}{3}\cot{[\mu(t-C_{2})]}\quad\textnormal{and}\quad q_{a}=3\sec^2{[\mu(t-C_{2})]}-1,
\end{equation}
and
\begin{equation}\label{2atau}
s = C_3\cos^{1/3}{[\mu(t-C_{2})]}
\end{equation}
yielding
\begin{equation}
H_{s}=-\frac{\mu}{3}\tan{[\mu(t-C_{2})]}\quad\textnormal{and}\quad q_{s}=3\csc^2{[\mu(t-C_{2})]}-1,
\end{equation}
with no restrictions on the range of $r$. We present plots of the scale factor in Fig. \ref{fig:sf3}, the deceleration parameter in Fig. \ref{fig:q3} of the external dimensions versus cosmic time $t$ for $n=3$. In Fig. \ref{fig:rho3} and Fig. \ref{fig:eos3}, we present plots of time $t$ evolution of the energy density and EoS parameter of the higher dimensional fluid. We note that both of the external and internal spaces oscillate with a period $P_{n=3}=\frac{\pi}{\mu}$ (time between the consecutive beginnings of the expansion of the external space). The constants $C_1$ and $C_3$ determine the oscillation amplitudes of the external and internal dimensions respectively.

\subsubsection{Subcase $r^2 - 27n\mu^2 < 0$}

\begin{figure}[h!]
     \begin{center}
        \subfigure[]{%
            \label{fig:sf3}
            \includegraphics[width=0.48\textwidth]{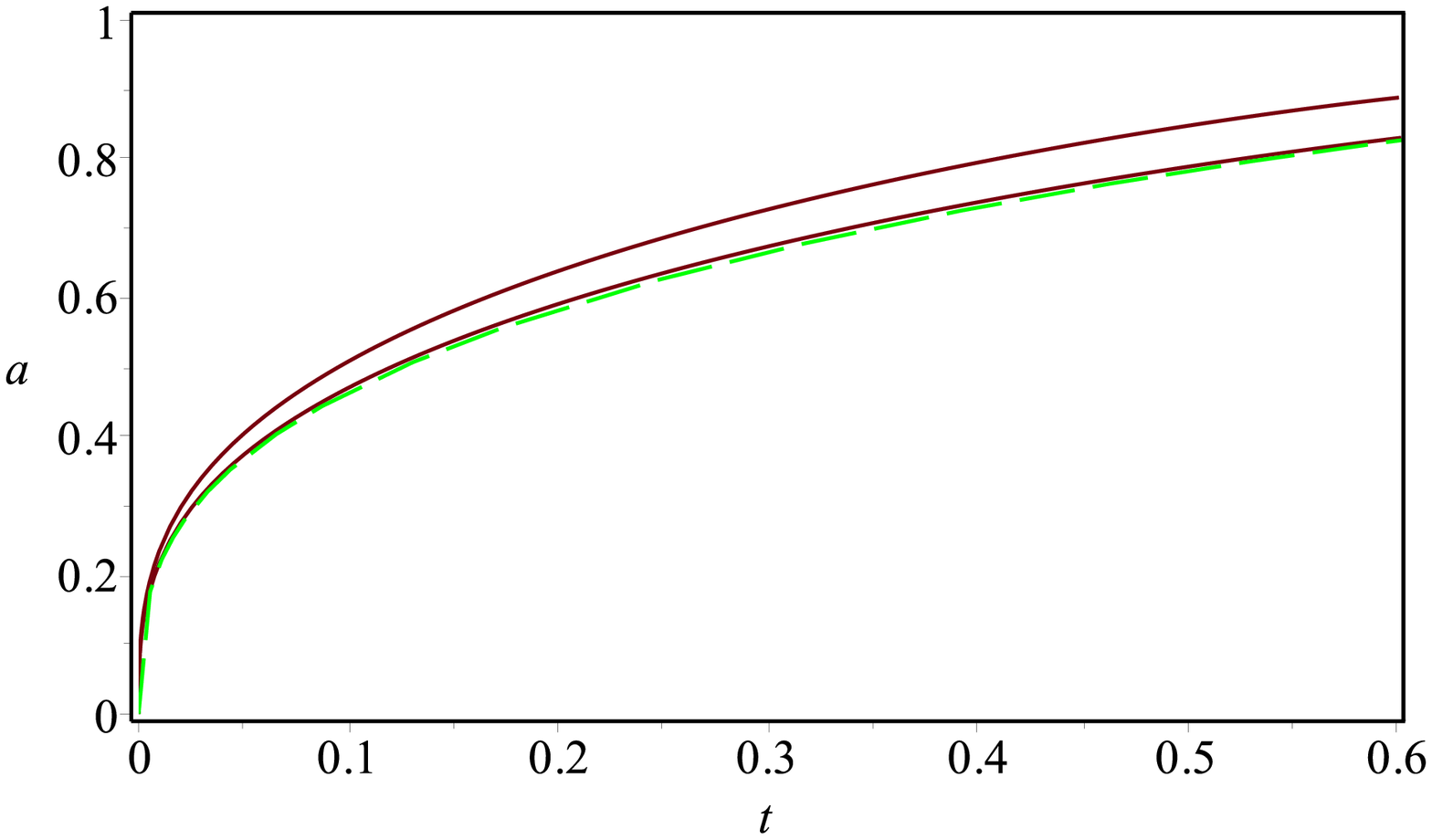}
        }%
        \subfigure[]{%
           \label{fig:q3}
           \includegraphics[width=0.48\textwidth]{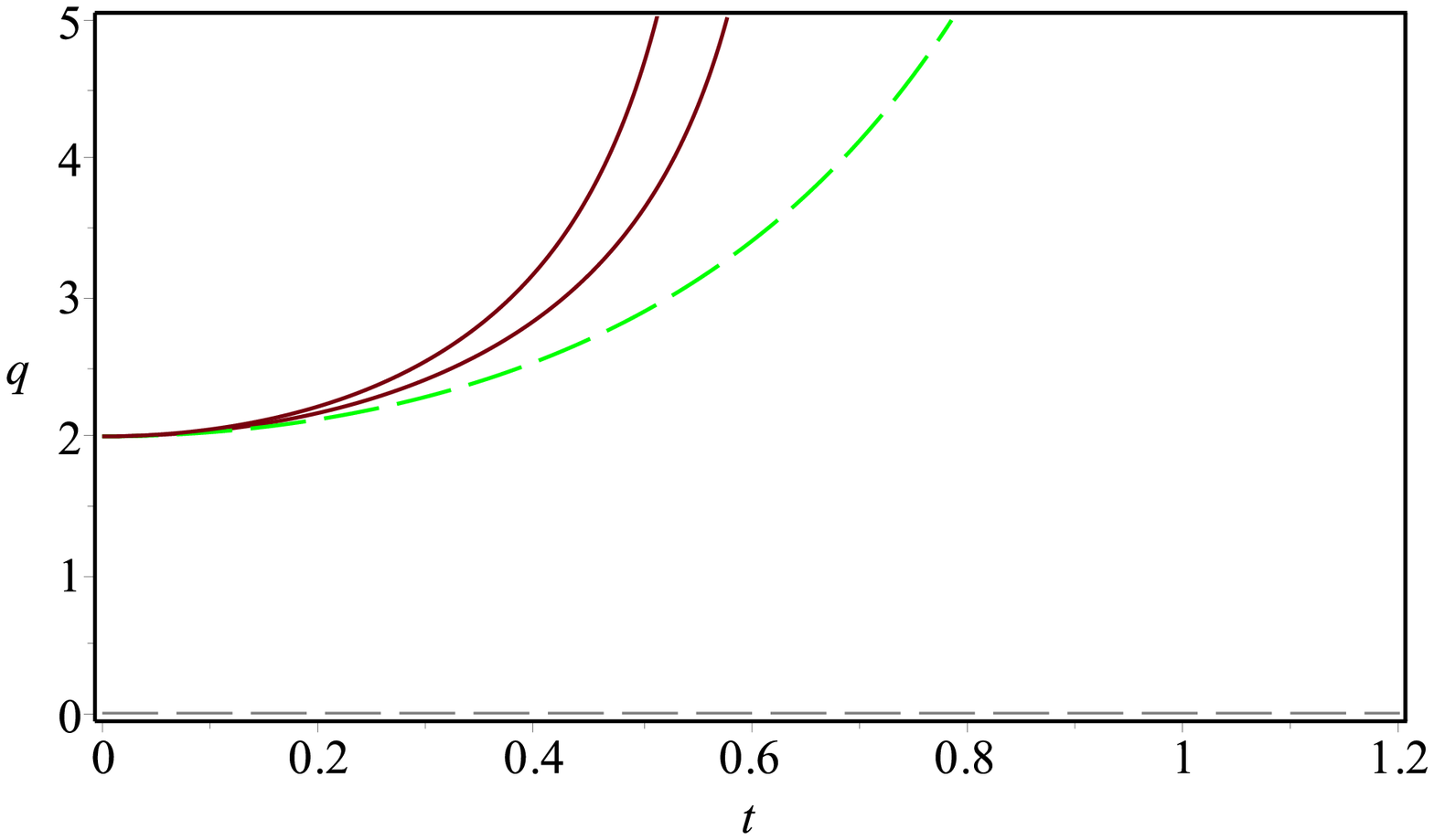}
        }

	\label{fig:OSrho2b}
	\subfigure[]{%
            \label{fig:rho3}
            \includegraphics[width=0.48\textwidth]{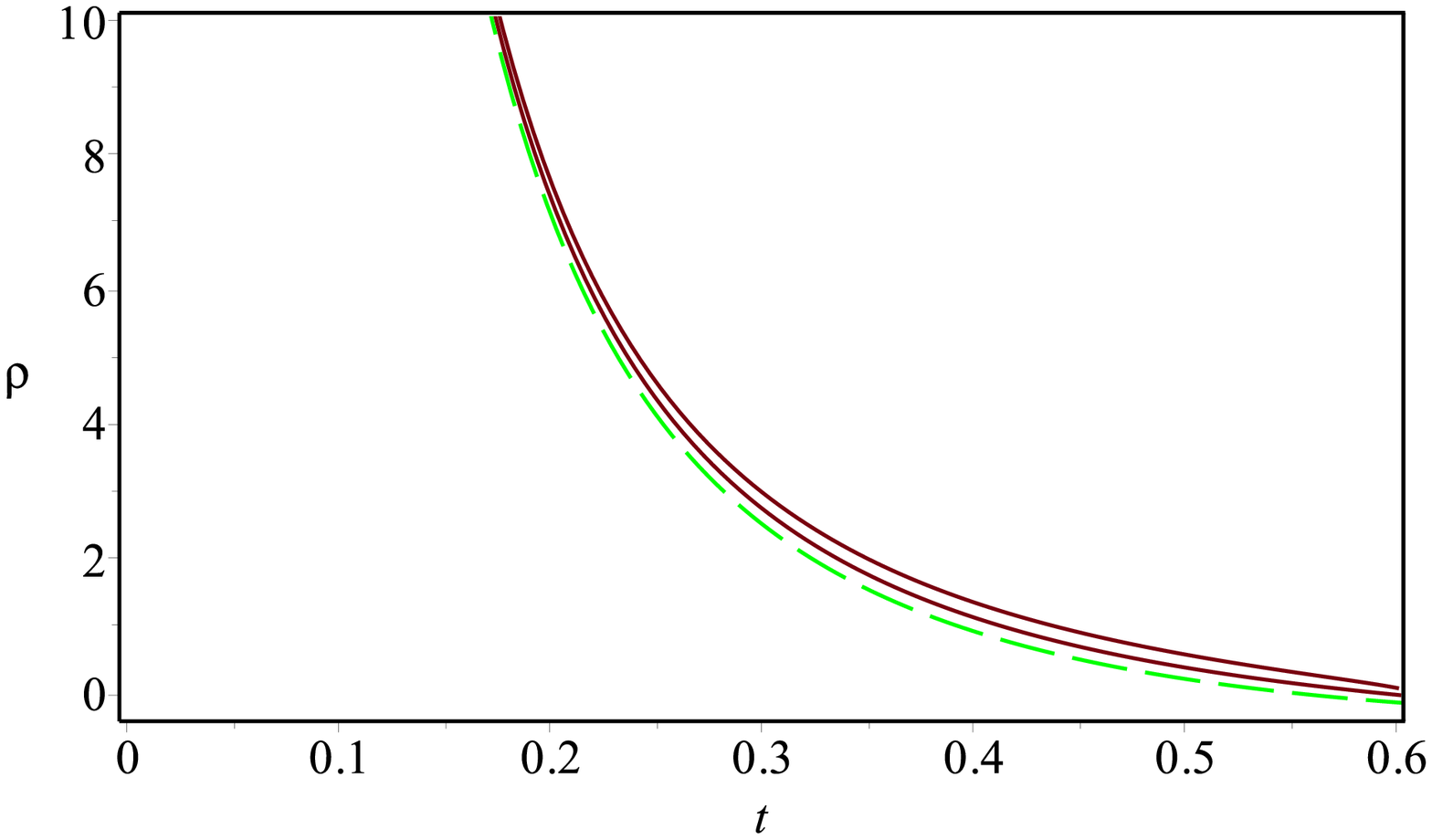}
        }%
        \subfigure[]{%
           \label{fig:eos3}
           \includegraphics[width=0.48\textwidth]{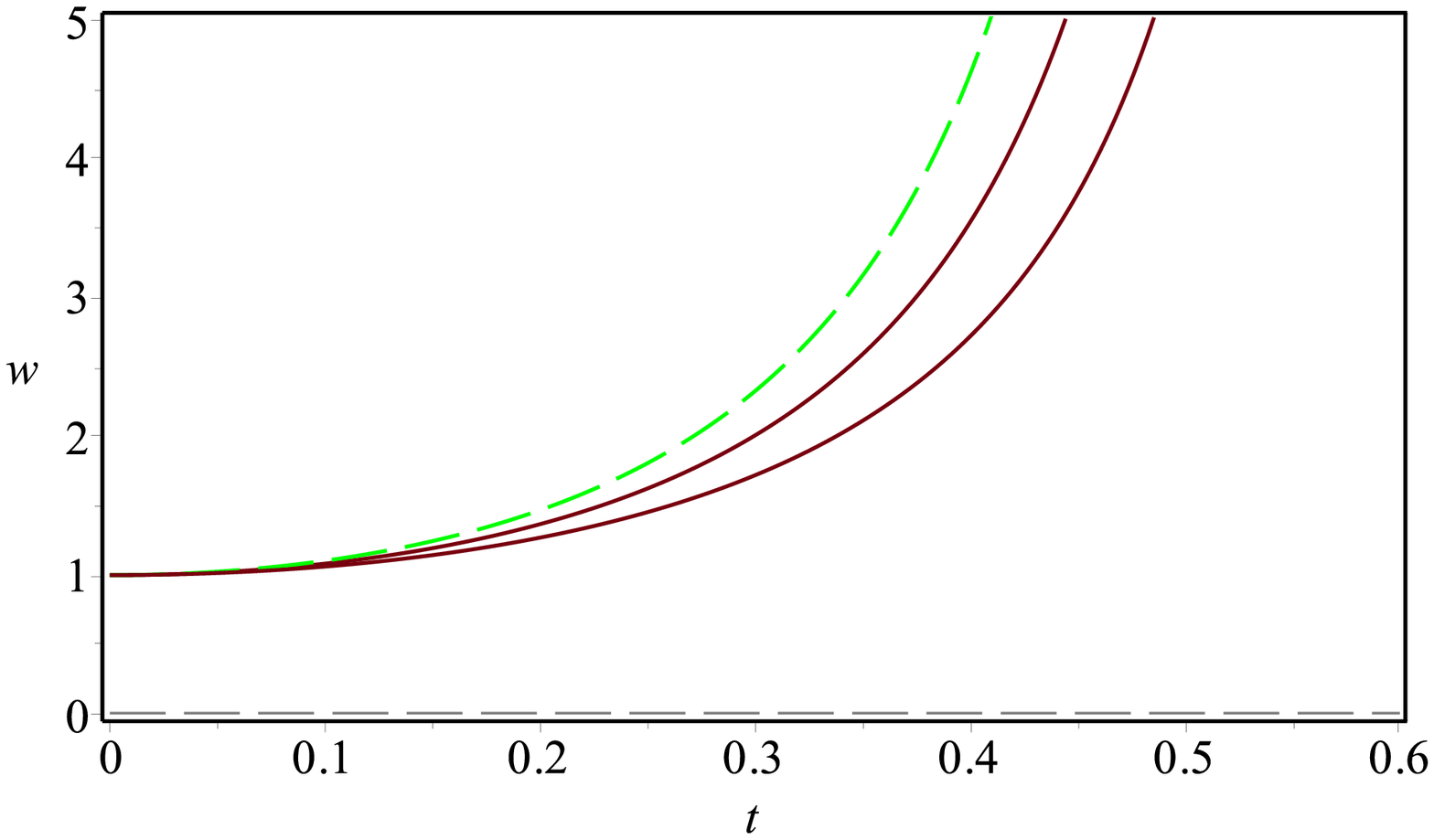}
        }
    \end{center}
	
	\caption{The evolution of some parameters of the model in cosmic time $t$ for $n=1$, $n=2$ and $n=3$, which give oscillating universes. The plots are given by choosing $\mu=1$ and hence their periods are as follows: $P_{n=1}=\frac{3}{4}\pi-\frac{\sqrt{3}}{2}\ln (2+\sqrt{3})$, $P_{n=2}=\frac{3}{5}\pi-\frac{\sqrt{6}}{10}\ln(5+2\sqrt{6})$ and $P_{n=3}=\pi$. The plots are given for the half period. (a) The scale factor of the external dimensions, $a$. (b) The deceleration parameter of the external dimensions, $q_{a}$. (c) The energy density of the higher dimensional fluid, $\rho$. (d) Equation of state (EoS) parameter of the higher dimensional fluid, $w$.}
\end{figure}

Within the range $r^2 - 27n\mu^2 < 0$ we have the following two cases that can be of interest in cosmology:

\subparagraph{(i) $n=1$ and $n=2$, with the ranges $r\in (-3n\mu,0)$ or $r\in (0,3n\mu)$:} If there are only one or two internal dimensions, i.e.  $n=1$ or $n=2$, then we have oscillating solutions within the ranges $r\in (-3n\mu,0)$ or $r\in (0,3n\mu)$. The oscillation periods are $P_{n=1}=\frac{3}{4\mu}\pi-\frac{\sqrt{3}}{2\mu}\ln (2+\sqrt{3})$, $P_{n=2}=\frac{3}{5\mu}\pi-\frac{\sqrt{6}}{10\mu}\ln(5+2\sqrt{6})$ for $n=1$ and $n=2$ respectively. One may check that $a\rightarrow0$, $H_{a}\rightarrow\infty$ and $q_{a}\rightarrow 2$ as $r\rightarrow0$. The Hubble parameters of the external and internal dimensions approach to non-zero constants as $r\rightarrow-3n\mu$, which give a delusive impression that the universe is dynamical at this limit. In fact, at this limit the expansion/contraction of the external/internal space ends and the contraction/expansion of the external/internal space starts: One may check that $\frac{{\rm d}t}{{\rm d}r}\rightarrow0$ as $r\rightarrow-3n\mu$, and that $\frac{{\rm d}a}{{\rm d}r}\rightarrow0$ and $\frac{{\rm d}s}{{\rm d}r}\rightarrow0$ as $r\rightarrow-3n\mu$, which means that the external space reaches its maximum size while the internal space reaches its minimum size at the limit $r\rightarrow-3n\mu$. The behavior of the model can also be investigated at the limit $r\rightarrow-3n\mu$ by considering the deceleration parameters $q_{a}$ and $q_{s}$ that are dimensionless and hence don't involve $t$ explicitly. One may check $q_{a}=\frac{\rm d}{{\rm d}t}\left(\frac{1}{H_{a}}\right)-1\rightarrow\infty$ and $q_{s}=\frac{\rm d}{{\rm d}t}\left(\frac{1}{H_{s}}\right)-1\rightarrow\infty$ as $r\rightarrow-3n\mu$, which also shows that the expansion/contraction of the external/internal space stops at the limit $r\rightarrow-3n\mu$. The energy density of the higher dimensional fluid $\rho$ starts with infinitely large values at $r=0$ both for $n=1$ and $n=2$, while, as $r\rightarrow-3n\mu$, it approaches zero and $-\frac{\mu^2}{3}$ for $n=1$ and $n=2$ respectively. The EoS parameter of the higher dimensional fluid starts with the value of $1$ and approaches $+\infty$ as $r\rightarrow-3n\mu$ both for $n=1$ and $n=2$. We present parametric plots of the scale factor in Fig. \ref{fig:sf3}, the deceleration parameter in Fig. \ref{fig:q3} of the external dimensions versus cosmic time $t$ for $n=1$ and $n=2$. In Fig. \ref{fig:rho3} and Fig. \ref{fig:eos3}, we present parametric plots of the cosmic time $t$ evolution of the energy density and EoS parameter of the higher dimensional fluid for $n=1$ and $n=2$.

\begin{figure}[h!]
     \begin{center}
        \subfigure[]{%
            \label{fig:sf4}
            \includegraphics[width=0.48\textwidth]{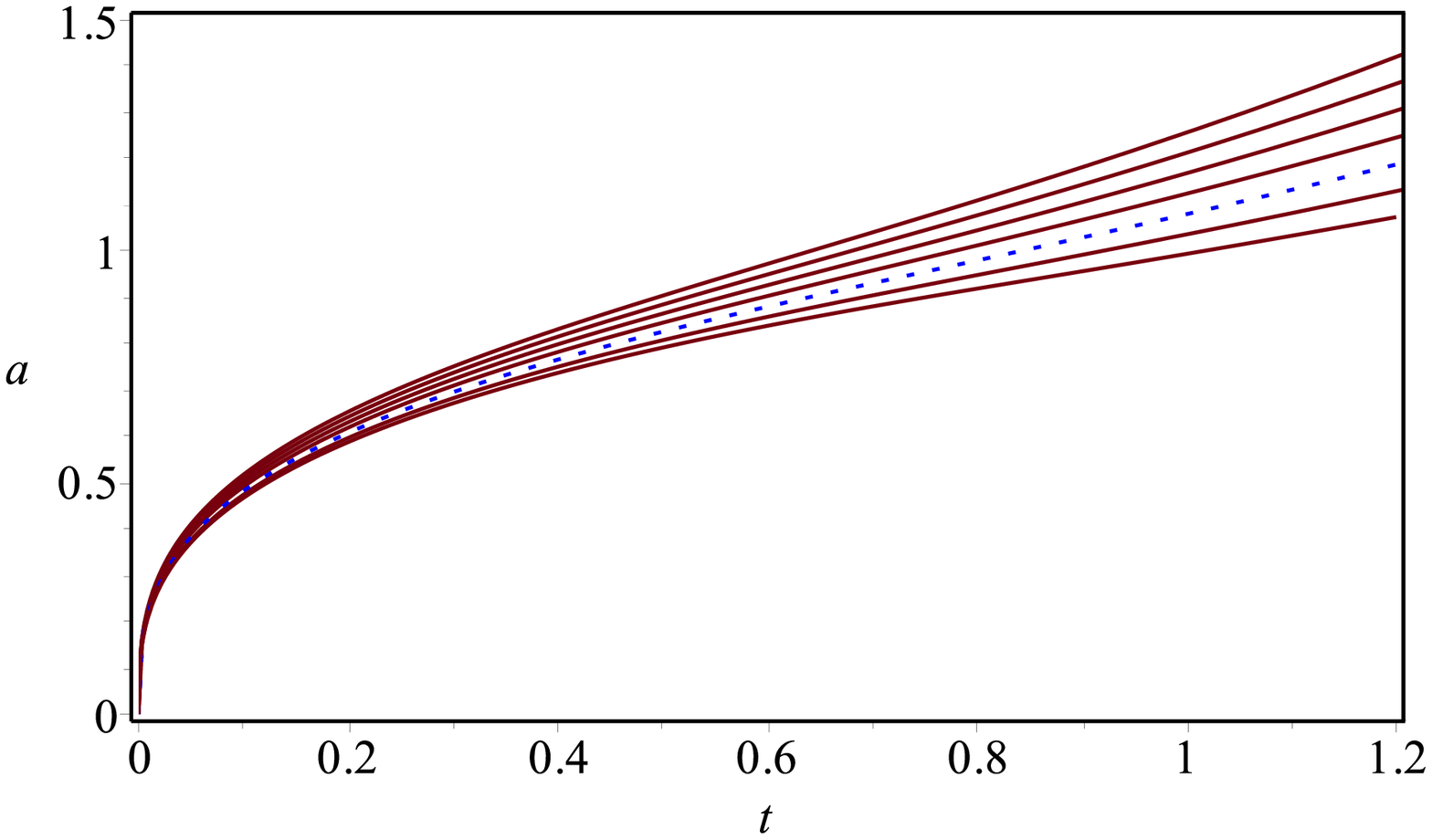}
        }%
        \subfigure[]{%
           \label{fig:q4}
           \includegraphics[width=0.48\textwidth]{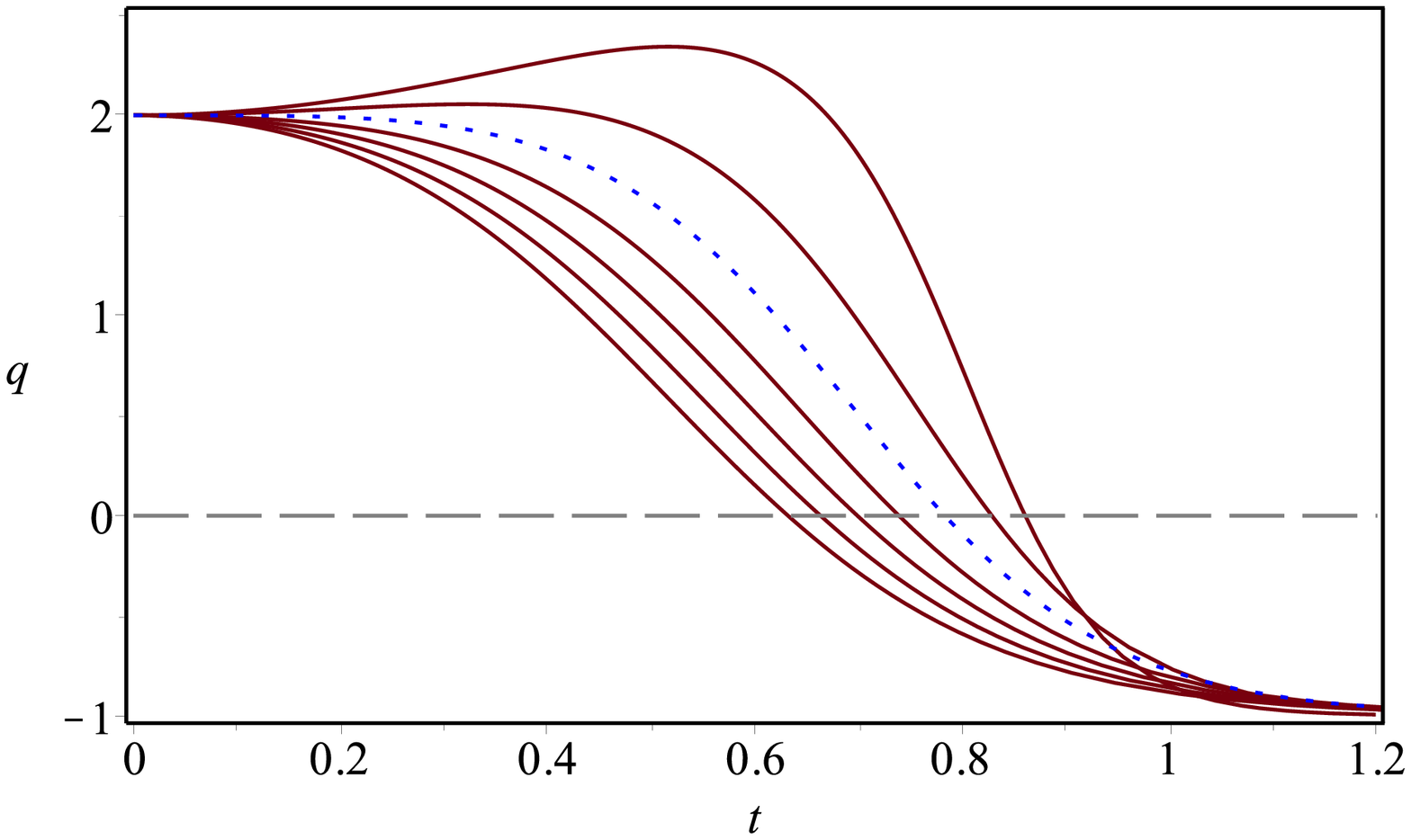}
        }

	\label{fig:rho2b}
	\subfigure[]{%
            \label{fig:rho4}
            \includegraphics[width=0.48\textwidth]{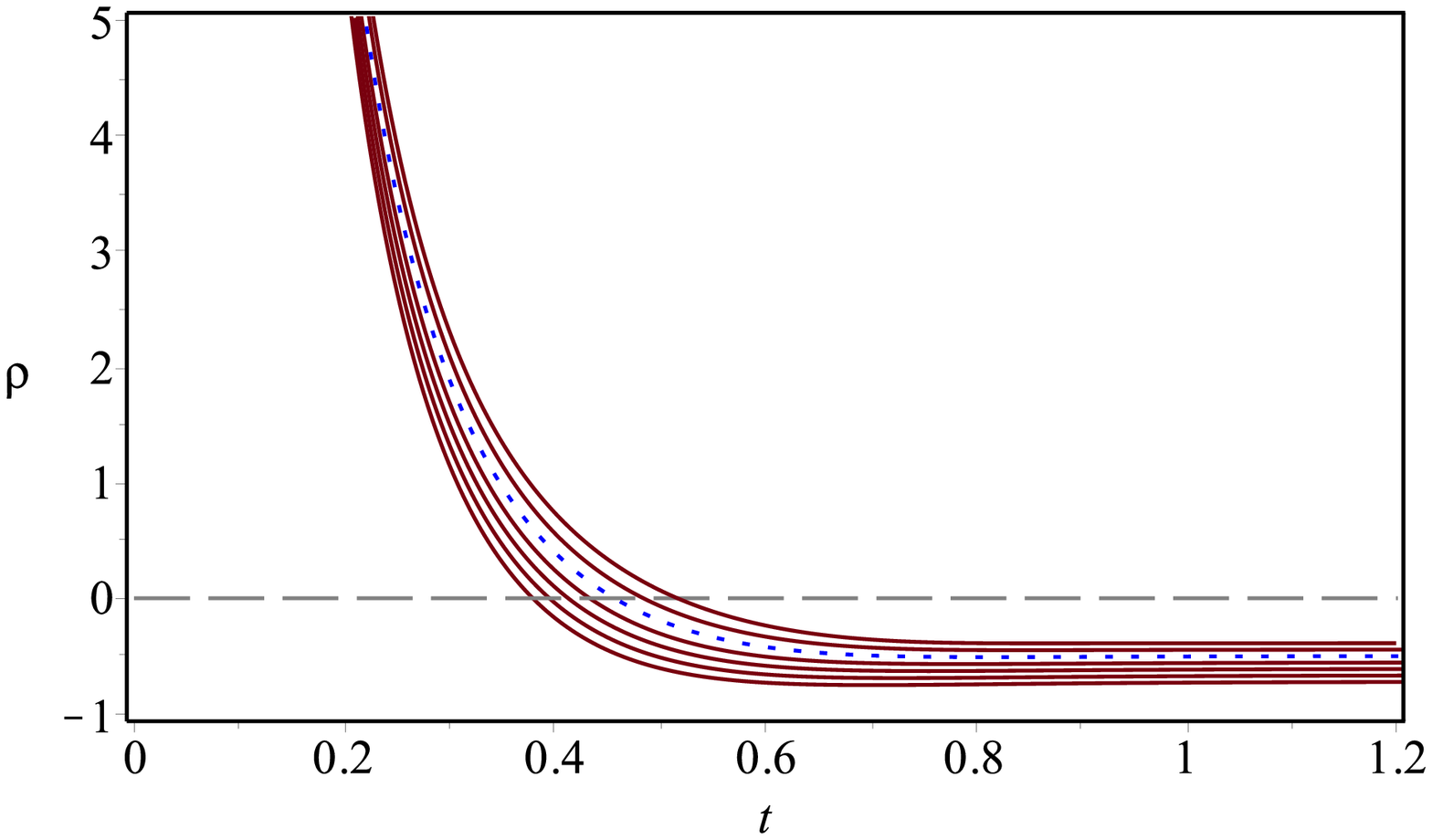}
        }%
        \subfigure[]{%
           \label{fig:eos4}
           \includegraphics[width=0.48\textwidth]{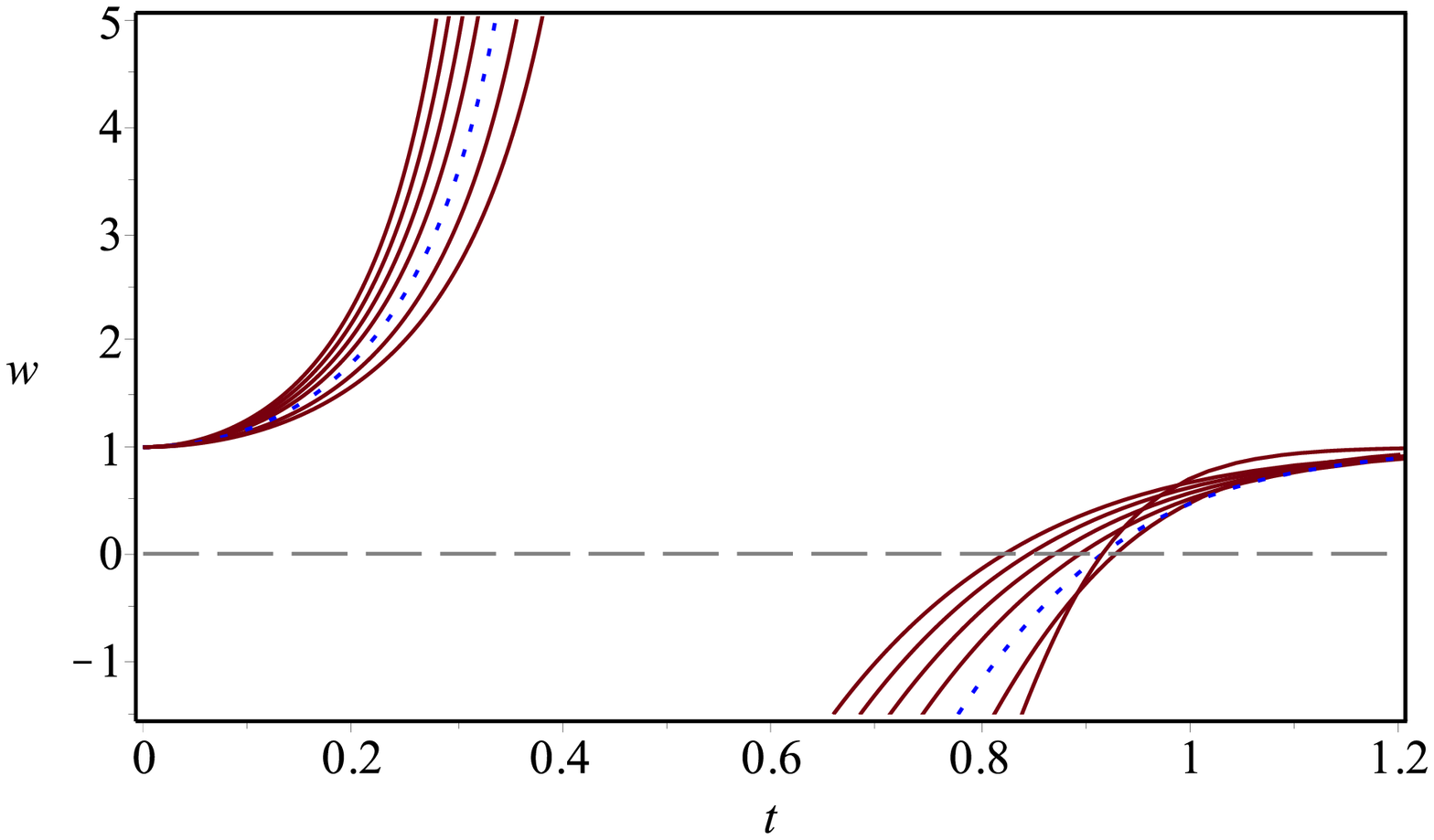}
        }
    \end{center}
	
	\caption{The evolution of some parameters of the model in cosmic time $t$ for $(4+n)$-dimensions. The plots are given from $n=4$ to $n=10$ by choosing $\mu=1$. The curves are in an order such that the dotted curves correspond to the case $n=6$. (a) The scale factor of the external dimensions, $a$. (b) The deceleration parameter of the external dimensions, $q_{a}$. (c) The energy density of the higher dimensional fluid, $\rho$. (d) Equation of state (EoS) parameter of the higher dimensional fluid, $w$.}
\end{figure}

\subparagraph{(ii) $n\geq4$, with the ranges $r\in (-3\sqrt{3n}\mu,0)$ or $r\in (0,3\sqrt{3n}\mu)$:} If there are more than three internal dimensions, i.e. $n\geq4$, then we have solutions, for which the external space exhibits type of behaviors similar to that of the $\Lambda$CDM model, within the ranges $r\in (-3\sqrt{3n}\mu,0)$ and $r\in (0,3\sqrt{3n}\mu)$. One may check that $a\rightarrow0$, $H_{a}\rightarrow\infty$ and $q_{a}\rightarrow 2$ as $r\rightarrow0$ and that $a\rightarrow\infty$, $H_{a}\rightarrow \frac{1}{3}\sqrt{\frac{n}{3}}\mu$ and $q_{a}\rightarrow -1$ as $r\rightarrow -3\sqrt{3n}\mu$. Accordingly, for all values of $n\geq4$, the external space starts with a decelerated expansion rate, enters into the accelerated expansion phase and eventually approaches exponential expansion. We note that $q'_{a}=0$ has two real roots for $n=4$ and $n=5$ ($r=0$ and $r=\pm \sqrt{n^2-n\sqrt{2n^2-6n}}$), while there is only one for $n\geq 6$ ($r=0$). According to this, only the cases $n\geq 6$ approaches exponential expansion ($q_{a}=-1$) monotonically. We present the parametric plots of the scale factor in Fig. \ref{fig:sf4} and the deceleration parameter in Fig. \ref{fig:q4} of the external dimensions versus cosmic time $t$ for $n=4$ to $n=10$. In Fig. \ref{fig:rho4} and Fig. \ref{fig:eos4}, we present the parametric plots of the cosmic time $t$ evolution of the energy density and EoS parameter of the higher dimensional fluid for $n=4$ to $n=10$. Note that the plots in these figures are depicted by redefining times as $t\rightarrow-t$ and $r\rightarrow -r$, so that the time parameter appears positive and the external space expands as $t$ increases in the figures. In these figures the dotted curves represent the case $n=6$. This solution possesses a noteworthy feature that should not be passed without mentioning. We note that the energy density of the higher dimensional fluid approaches a non-zero constant and its EoS parameter approaches the value of $1$ as $a\rightarrow\infty$. Considering the conservation of the energy-momentum tensor for a minimally coupling energy source, i.e. $\dot{\rho}+(3H_{a}+nH_{s})(1+w)\rho=0$, constant energy density is possible only if the EoS parameter is equal $-1$ or if the volume is constant $V=a^3s^n={\rm constant}$, i.e., $3H_{a}+nH_{s}=0$. Indeed using the Hubble parameters of the external and internal dimensions from \eref{sonha} and \eref{sonhs} we find that
\begin{equation}
\frac{\dot{V}}{V}=3H_{a}+nH_{s}=-\frac{3n\mu^2}{r}+\frac{r}{9}
\end{equation}
and that $H_a\rightarrow\frac{1}{3}\sqrt{\frac{n}{3}}\mu$, $H_s\rightarrow-\frac{\mu}{\sqrt{3n}}$ and $\frac{\dot{V}}{V}\rightarrow0$ as $r\rightarrow -3\sqrt{3n}\mu$. Hence the total volume of the universe approaches a finite size as $t\rightarrow\infty$, while the external(internal) space keeps on expanding(contracting) forever. Accordingly, in this model, the universe approaches a higher dimensional steady state universe, that is characterized by dynamical external and internal spaces having a constant $(3+n)$-dimensional volume and a constant mean energy density in $(3+n)$ dimensions \cite{Akarsu13b,Akarsu13c}.

\subsubsection{Subcase $r^2 - 27n\mu^2 > 0$}

\begin{figure}[h!]
     \begin{center}
        \subfigure[]{%
            \label{fig:qlast1}
            \includegraphics[width=0.31\textwidth]{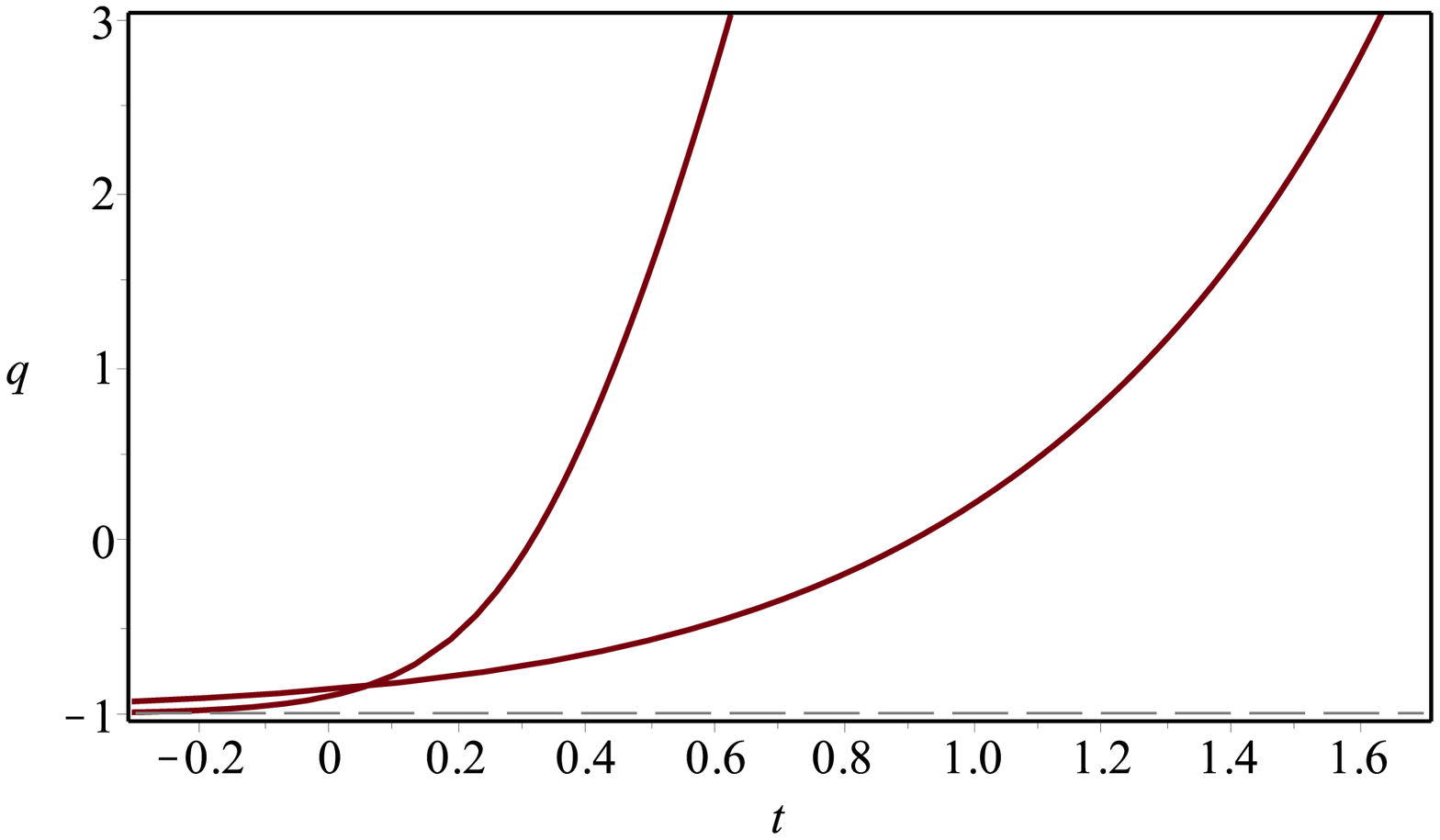}
        }
        \subfigure[]{%
           \label{fig:qlast2}
           \includegraphics[width=0.31\textwidth]{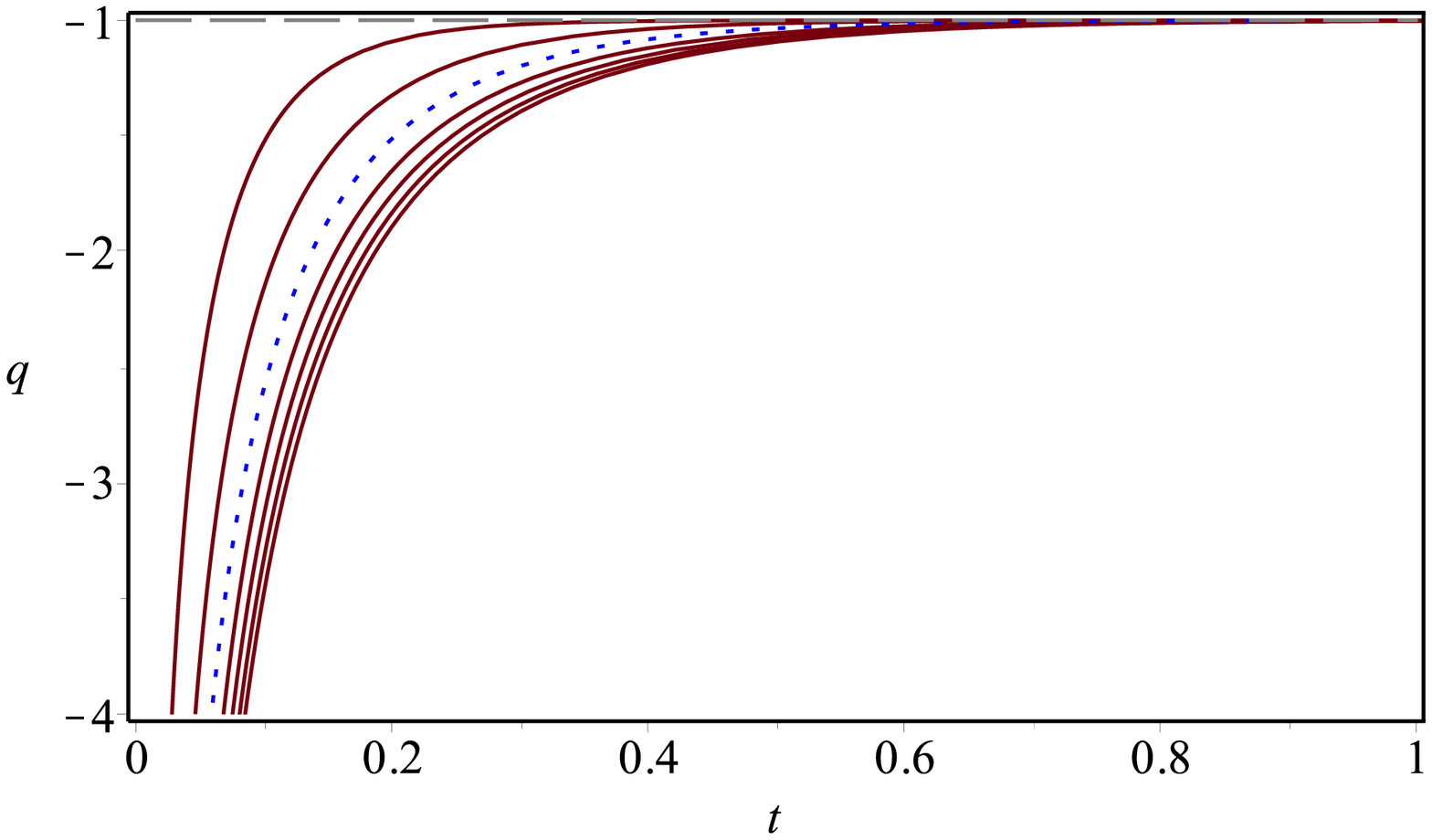}
        }
		\subfigure[]{%
            \label{fig:qlast3}
            \includegraphics[width=0.31\textwidth]{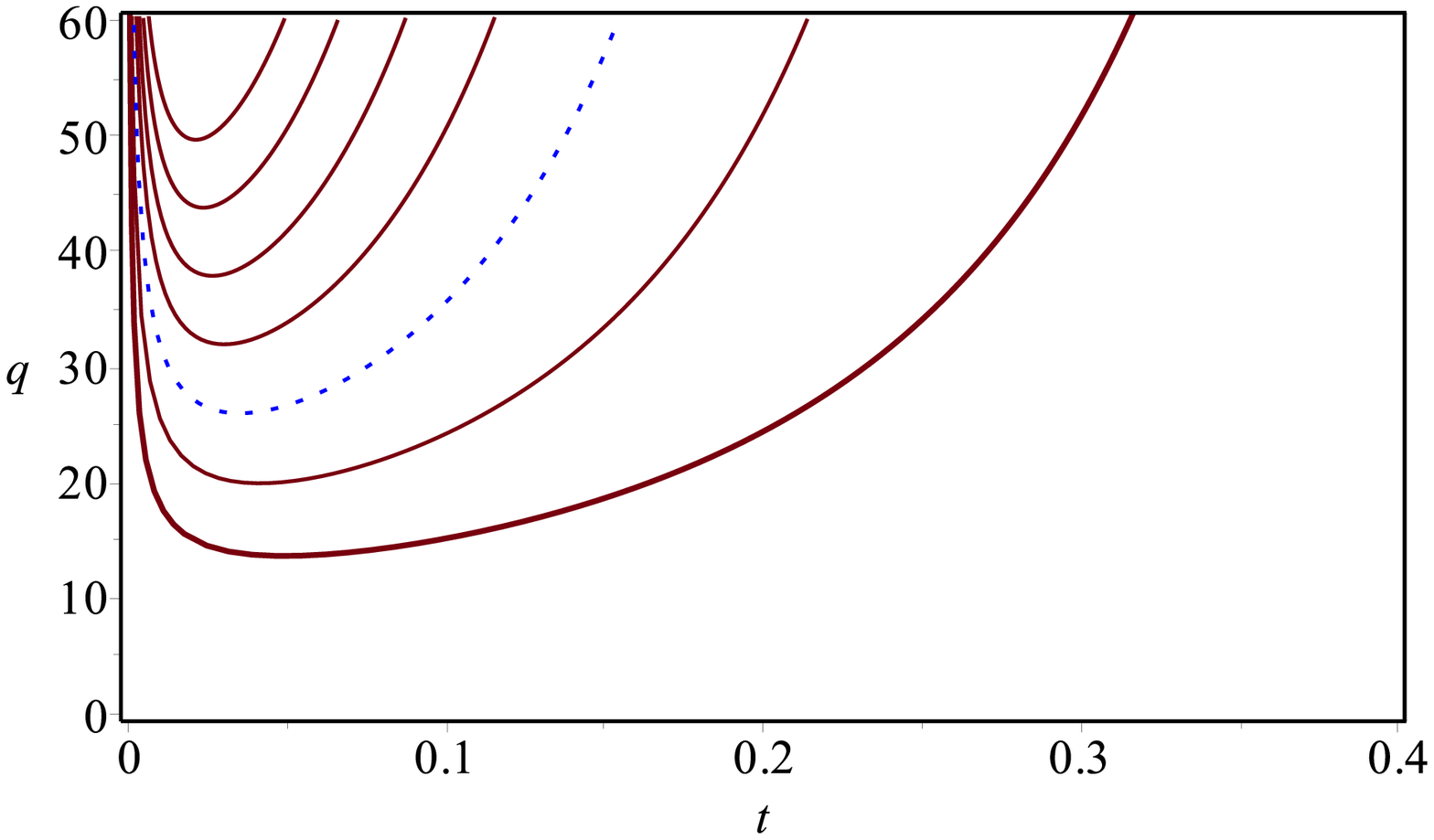}
        }

    \end{center}
	
	\caption{The evolution of the deceleration parameter of the external space in cosmic time $t$ for three different solutions in the case $r^2 - 27n\mu^2 > 0$. The plots are given by choosing $\mu=1$. We redefine the cosmic time as $t\rightarrow-t$ for convenience. (a) The cases $n=1$ and $n=2$ in the range $r\in (3\sqrt{3n}\mu,+\infty)$. (b) The cases $n\geq4$ in the range $r\in (3\sqrt{3n}\mu,3n\mu)$. The curves are given in an order such that the dotted curve correspond to case $n=6$. (c) The cases $n\geq4$ in the range $r\in (3n\mu,+\infty)$.  The curves are given in an order such that the dotted curve correspond to case $n=6$. }
\end{figure}

As it is listed above in this section for $r^2 - 27n\mu^2 > 0$ we have three different set of solutions when we consider the evolution of the universe in terms of cosmic time $t$. In comparison with the cases discussed above, the behavior of the external space in all these three cases is less interesting from the cosmological point of view. Therefore, for the sake of brevity, our discussion in this section will be confined by the kinematics of the external space.
\subparagraph{(i) $n=1$ and $n=2$, with the ranges $r\in (-\infty,-3\sqrt{3n}\mu)$ or $r\in (3\sqrt{3n}\mu,+\infty)$:} We note that $a\rightarrow0$, $q\rightarrow-1$ as $r\rightarrow 3\sqrt{3n}\mu$ and $a\rightarrow a_{\rm max}={\rm const.}\neq0$, $q\rightarrow\infty$ as $r\rightarrow \infty$. The cosmic time $t$, on the other hand, ranges from $t=\infty$ to $t=-\frac{3\pi}{(3+n)\mu}+C_2$ as $a$ ranges from $a=0$ to $a={\rm const.}\neq0$. According to this the external space starts to contract from its maximum size and keeps on contraction forever by approaching exponential contraction at infinite future. We can interpret this result in another way. If we redefine time as $t\rightarrow-t$, then we have an expanding external space that approaches to a finite maximum size coming from a de Sitter expansion phase in the infinite past. We present deceleration parameter of the external space $q_{a}$ versus cosmic time $t$ in fig. \ref{fig:qlast1} for the latter interpretation.
\subparagraph{(ii) $n\geq4$, with ranges $r\in (-3n\mu,-3\sqrt{3n}\mu)$ or $r\in (3\sqrt{3n}\mu,3n\mu)$:} We note that $a\rightarrow\infty$, $q\rightarrow-1$ as $r\rightarrow 3\sqrt{3n}\mu$ and $a\rightarrow a_{\rm min}={\rm const.}\neq0$, $q\rightarrow -\infty$ as $r\rightarrow 3n\mu$. The cosmic time $t$, on the other hand, ranges from $t={\rm const.}$ to $t=-\infty$ as $a$ ranges from $a=a_{\rm min}$ to $a=\infty$. Accordingly, redefining the cosmic time as $t\rightarrow-t$, we find that the external space starts expansion from its minimum size and then keeps on expanding forever by approaching an exponential expansion. We present deceleration parameter of the external space $q_{a}$ versus cosmic time $t$ in fig. \ref{fig:qlast2}.
\subparagraph{(iii) $n\geq4$, with ranges $r\in (-\infty,-3n\mu)$ or $r\in (3n\mu,+\infty)$:} We note that both the cosmic time $t$ and the scale factor factor of the external space $a$ ranges between two different finite values between the limits $r\rightarrow3n\mu$ and $r\rightarrow+\infty$, while the deceleration parameter of the external space $q_{a}$ is identically becomes infinitely large at these limits. All these indicate the oscillating behavior of the universe in this case, i.e., the external space oscillates between its finite extremums. We present deceleration parameter of the external space $q_{a}$ versus cosmic time $t$ in fig. \ref{fig:qlast3}.

\section{Concluding Remarks}

In this paper we studied the general solution of a higher dimensional cosmological model \cite{AkDe} that is characterized by a single real parameter $\lambda$, the product of the Hubble parameters of the internal and external spaces,  that correlates and controls the dynamical evolution of the cosmology. In the original study the solutions of the field equations were given only for the particular case for which an explicit solution exists, that is when the number of internal dimensions is $n = 3$, and  $\lambda > 0$ only. In the present paper, on the other hand, we obtain the general solution of the system for arbitrary values of $n=1,2,3,...$ and  $\lambda$, expressed  in parametric form with the help of Lie symmetry properties. We also provide explicit analytic solutions of the system in terms of the cosmic time $t$ for special cases i) $\lambda=0$ for arbitrary values of $n$ and ii) $\lambda\neq0$ for $n=3$.

\medskip

\noindent 1. We show that depending on  the range of $r$, the values $n=1,2,3,\dots$ may take, as well as the sign of $\lambda$ our parametric solution may lead to very different types of cosmological evolution in $t$. Such a diversity of cosmological dynamics depending on the parameter values wouldn't be
apparent if we  looked for analytic solutions in terms of cosmic time $t$ only. In this sense our parametric solution is important.

\medskip

\noindent 2. Even though we cannot express our general solution analytically in terms of $t$, we are able to generate plots for all $n=1,2,3,\dots$  of physically relevant quantities such as the scale factors, Hubble and deceleration parameters of both the internal and external spaces as functions of $t$.
The energy density, pressure and EoS in each case are also plotted as functions of $t$. It is remarkable that the cases $n=3$ and $n=6$ stand out among others as critical dimensions at which qualitative changes in the evolutionary behavior of the universe occur.

\medskip

3. In this paper we obtained different cosmological models depending on the sign of $\lambda$, the number of internal dimensions $n$ and the range of the parameter $r$. To be concise we haven't discussed some aspects of the models, such as the 4-dimensional effective universe as it was done for the particular case $\lambda>0$ and $n=3$ in the original study \cite{AkDe}. However, it might be useful to comment on the 4-dimensional effective gravitational coupling $\tilde{\kappa}$ as it is dynamical in our solutions. It is well known that  4-dimensional effective gravitational coupling is inversely proportional to the volume of the internal space $V_{\rm int}=s^n$ while it is proportional with the $(1+3+n)$-dimensional gravitational coupling $\kappa$ \cite{Dvali99,Uzan11}. Accordingly, as $\kappa$ is a constant in our model, the rate of change of the 4-dimensional effective gravitational coupling simply reads
\begin{equation}
\frac{\dot{\tilde{\kappa}}}{\tilde{\kappa}}=-n\frac{\dot{s}}{s}=-\frac{n\lambda a}{\dot{a}}.
\end{equation}
Since $\dot{s}/s=r/(9n)$, this also leads to a simple relationship between the parameter $r$ and the rate of change of 4-dimensional effective gravitational coupling as
\begin{equation}
\label{varG}
\frac{\dot{\tilde{\kappa}}}{\tilde{\kappa}}=-\frac{r}{9}.
\end{equation}
Using the above relationship in \eqref{dr/dtau}, we find that the differential equation that describes the rate of change of the 4-dimensional effective gravitational coupling is as follows:
\begin{equation}
    \frac{{\rm d}}{{\rm d}t} \left(\frac{\dot{\tilde{\kappa}}}{\tilde{\kappa}} \right) = \frac{\left(9(\frac{\dot{\tilde{\kappa}}}{\tilde{\kappa}})^2+3n\lambda\right)\left(9(\frac{\dot{\tilde{\kappa}}}{\tilde{\kappa}})^2-n^2\lambda\right)}{9\left(9(\frac{\dot{\tilde{\kappa}}}{\tilde{\kappa}})^2+n^2\lambda\right)}.
\end{equation}
We would like to note that \eqref{varG} also reveals the physical meaning of $r$ which has been utilized for obtaining the exact parametric solution of the model and then lets us give various cosmological scenarios in cosmic time $t$ depending on the considered range of $r$ that we gave in different sections. Hence the ranges of $r$ that correspond to different models in cosmic time $t$ amount to setting the ranges for the rate of change of 4-dimensional effective gravitational coupling in cosmic time. Accordingly we can utilize the ranges of the solutions to have an idea about the $\dot{\tilde{\kappa}}/\tilde{\kappa}$ in our models. The most promising two models among all are those where the value of the deceleration parameter of the external space evolves from 2 to -1 as the external space expands: the case $\lambda>0$ and $0\leq r \leq 3n\sqrt{\lambda}$ (given in Section (3.1.1)) and the case $\lambda=-\mu^2>0$ and $- 3\sqrt{3n}\mu \leq r \leq 0$ for $n\geq 4$ (case (ii) given in Section (3.2.2)). In these two cases the $\dot{\tilde{\kappa}}/\tilde{\kappa}$ is null when $a=0$ ($r=0$). As the external space expands, in the former case $\tilde{\kappa}$ decreases and $\dot{\tilde{\kappa}}/\tilde{\kappa}\rightarrow -n\sqrt{\lambda}/3$ as $a\rightarrow\infty$. In the latter one $\tilde{\kappa}$ increases and $\dot{\tilde{\kappa}}/\tilde{\kappa}\rightarrow \sqrt{\frac{n}{3}}\mu$ as $a\rightarrow\infty$. We note that in neither of the cases the magnitude of $\dot{\tilde{\kappa}}/\tilde{\kappa}$ does not grow indefinitely and hence may be set to sufficiently small values that couldn't be detected through the history of the universe. On the other hand, we give also solutions where the magnitude of $r$ and hence that of $\dot{\tilde{\kappa}}/\tilde{\kappa}$ tends to infinitely large values. It is interesting that the evolution of the external space in these cases is also drastically different from the expansion history of the universe we observe. The detailed discussion of the observational constraints on all of these models is out of the scope of this paper. However, a detailed discussion on $\dot{\tilde{\kappa}}/\tilde{\kappa}$ for the particular case $\lambda>0$ and $n=3$ can be found in \cite{AkDe}. There the average value of $\dot{\tilde{\kappa}}/\tilde{\kappa}$ from $t=0$ to the present age of the universe $13.7$ (Gyr) was calculated as $\sim10^{-11}\,{\rm yr}^{-1}$ and it was found that $\dot{\tilde{\kappa}}/\tilde{\kappa}\sim -10^{-25}\,{\rm yr}^{-1}$ for the time scale $\sim 10^{2}\,{\rm s}$ which is the time scale when the primordial nucleosynthesis took place that sets the most severe constraints on $\dot{\tilde{\kappa}}/\tilde{\kappa}$ as $\sim 10^{-12}\,{\rm yr}^{-1}$.

\medskip

\noindent 4. The last point we wish to emphasize concerns the sign of our correlation parameter $\lambda$. If we consider only the accelerated expansion of the external space (as the observations dictate) then the internal space contracts for $\lambda < 0$ while it also expands for $\lambda >0$. The picture in the first case is typical for many higher dimensional cosmological models that are discussed in the literature. It often happens in such models that as  $t$ gets very large $\rho$ and/or $p$ as well as the scale factor of the internal space may hit singularities. We also observe in this case solutions where the energy density of the higher dimensional fluid $\rho$ goes negative. One may try to shift it to positive values by introducing a negative vacuum energy, which is the only allowed negative energy source satisfying the dominant energy condition \cite{Carroll01} and often appears in unified theories such as string theory \cite{Sahni00,Kachru03,Nobbenhuis06}. On the other hand, the choice $\lambda >0$, that is the case in which the internal space expands too, discussed here is not typical as far as we know and allows us to avoid running into such difficulties. Moreover, the notion of an expanding internal space may be tempting in the context of hierarchy problem (See \cite{Dvali99} and references therein for hierarchy problem). If the fields of standard theory of particles are confined to our $3$-dimensional space by a suitable mechanism, then the expansion of the internal space will dynamically reduce only the $4$-dimensional effective gravitational coupling. Provided that there are sufficiently large number of internal dimensions, the volume of the internal space may increase to a size large enough to explain the weakness of gravity relative to other fundamental forces but yet the internal dimensions remain at an unobservable size.

\medskip

\noindent We thus demonstrated the viability of a new class of higher dimensional cosmological models for which both external and internal
dimensions are  at comparably small scales during the early stages of their evolution and at later stages the internal
dimensions expand at a much slower rate than those of the external space
and remain unobservable.

\bigskip

\begin{center}
\textbf{Acknowledgments}
\end{center}
\"{O}.A. acknowledges the support by T\"{U}B{\.I}TAK Research Fellowship for Post-Doctoral Researchers (2218). \"{O}.A. and T.D. acknowledge the support from Ko\c{c} University. N. Kat{\i}rc{\i} thanks  Bo\u{g}azi\c{c}i University for the financial support provided by the Scientific Research Fund with BAP project no: 7128. The research of M.B. Sheftel was supported in part by the research grant from  Bo\u{g}azi\c{c}i  University Scientific Research Fund (BAP), research project No. 6324.

\end{document}